\documentclass[%
 aip,
 amsmath,amssymb,
reprint,%
]{revtex4-1}
\usepackage[graphicx]{realboxes}
\usepackage{graphicx}
\usepackage{comment}
\usepackage{xcolor}
\usepackage{booktabs, multirow}
\usepackage{dcolumn}
\usepackage{bm}

\usepackage[utf8]{inputenc}
\usepackage[T1]{fontenc}
\usepackage{mathptmx}
\usepackage{etoolbox}
\usepackage[version=3]{mhchem} 

\makeatletter
\def\@email#1#2{%
 \endgroup
 \patchcmd{\titleblock@produce}
  {\frontmatter@RRAPformat}
  {\frontmatter@RRAPformat{\produce@RRAP{*#1\href{mailto:#2}{#2}}}\frontmatter@RRAPformat}
  {}{}
}%
\makeatother
\begin{document}

\preprint{AIP/123-QED}

\title[Probabilistic and Entropy Modeling]{Probabilistic and Maximum Entropy Modeling of Chemical Reaction Systems: Characteristics and Comparisons to Mass Action Kinetic Models}

\author{William R. Cannon}
\affiliation{Physical and Computational Sciences Directorate, Pacific Northwest National Laboratory, Richland, WA 99352, USA}%
\affiliation{Department of Mathematics, University of California, Riverside, CA, 92505, USA}
\affiliation{Center for Quantitative Modeling in Biology, University of California Riverside, Riverside, CA, 92505, USA}
\email[Author to whom correspondence should be addressed:]{William.Cannon@pnnl.gov}

\author{Samuel Britton}%
 \email{samuelryanbritton@gmail.com}
\affiliation{Department of Mathematics, University of California, Riverside, CA, 92505, USA}
\affiliation{Center for Quantitative Modeling in Biology, University of California Riverside, Riverside, CA, 92505, USA}

\author{Mikahl Banwarth-Kuhn}%
 \email{mikahl.banwarthkuhn@csueastbay.edu}
\affiliation{Department of Mathematics,  California State University East Bay, Hayward, CA 94542, USA}
\affiliation{Center for Quantitative Modeling in Biology, University of California Riverside, Riverside, CA, 92505, USA}
\altaffiliation[Currently at ]{Department of Mathematics,  California State University East Bay}

\author{Mark Alber}%
 \email{malber@ucr.edu.}
\affiliation{Department of Mathematics, University of California, Riverside, CA, 92505, USA}
\affiliation{Center for Quantitative Modeling in Biology, University of California Riverside, Riverside, CA, 92505, USA}

\date{\today}

\begin{abstract}
We demonstrate and characterize a first-principles approach to modeling the mass action dynamics of metabolism. Starting from a basic definition of entropy expressed as a multinomial probability density using Boltzmann probabilities with standard chemical potentials, we derive and compare the free energy dissipation and the entropy production rates.  We express the relation between the entropy production and the chemical master equation for modeling metabolism, which unifies chemical kinetics and chemical thermodynamics. Subsequent implementation of a maximum entropy model for systems of coupled reactions is accomplished by using an approximation to the Marcelin equation for mass action kinetics. Because prediction uncertainty with respect to parameter variability is frequently a concern with mass action models utilizing rate constants, we compare and contrast the maximum entropy model, which has its own set of rate parameters, to a population of standard mass action models in which the rate constants are randomly chosen. We show that a maximum entropy model is characterized by a high probability of free energy dissipation rate, and likewise entropy production rate, relative to other models. We then characterize the variability of the maximum entropy model predictions with respect to uncertainties in parameters (standard free energies of formation) and with respect to ionic strengths typically found in a cell. 
\end{abstract}

\maketitle

\section{\label{sec:level1}Introduction}

The maximization of entropy has been alluded to historically or directly stated as the goal or an emergent property of biological systems by both physicists and ecologists \cite{Dewar2010,Lotka1922,Odum1955,Prigogine1978,Schrodinger1945,Dyke2010,Harte2011,Kleidon2010,Meysman2009,Unrean2011,Vallino2009}. Yet, the concept has been underdeveloped regarding application to systems such as metabolism, and because of the abstract nature of the concept, it has gained insufficient recognition as an operational principle among microbiologists and cell biologists. 
There are several issues regarding the application of the principle of maximum entropy to biological systems. 

First and foremost, it is not always clear what is meant by ‘entropy’. The term has been associated with several mathematical descriptions, not all of which are equivalent, for example, even as they are used in physics and chemistry. The works of Ge and Qian \cite{Ge2010}, Presse, \textit{et al.} \cite{Presse2013}, Watchel, \textit{et al} \cite{watchel_2022}, Seifert \cite{Seifert_review_2019} and Cannon \cite{Cannon2014a} provide foundations to build on, in this regard. However, 
even given a rigorous thermodynamic definition of entropy, it is not clear how the maximum entropy concept would be applied to modeling a specific system such as metabolism, metabolic regulation or protein expression dynamics and its regulation. 

Second, the concept can seem simplistic to those not familiar with the depth of the theory and its implications, leading to antagonism towards the idea that cells are thermodynamic machines. Although the idea is simple at a high level of abstraction, its application is not simple and does not imply that all we need is to change our perspective and we will understand biology. However, applying the principle can uncover and aid in the understanding of emergence and biological complexity.

Third, when applied to dynamical systems, what is often meant is that the entropy (or free energy) produced is maximized (or minimized, respectively). But the concept of maximum entropy in statistical thermodynamics goes beyond simple predictive statements about a system moving to the most probable state, \textcolor{black}{in agreement with the second law of classical thermodynamics,} to include statistical inference.
In fact, when the term maximum entropy or maximum path entropy is used in modeling dynamical systems, the term maximum caliber is often used – the application of maximum entropy concepts to dynamical systems with the use of constraints. When applied using constraints, the maximum caliber principle attempts to maximize a statistical entropy function or empirical distribution of unobserved variables while constraining the observed variables to their experimentally measured values \cite{Jaynes1985,Presse2013}. The constraints might be those due to which chemical transformations (reactions) are possible, or a set of concentrations that are experimentally observed.

Applying sophisticated thermodynamics principles to a biological system is necessary but not sufficient to understand how cells function. There are many emergent processes and structures that occur in a cell that we do not understand sufficiently to infer with thermodynamics. Moreover, the functional capabilities of a cell – it’s set of metabolic reactions, gene complement, regulation, etc. – are specifically tuned to its environmental niche, and, in principle, represent a local thermodynamic optimum, arrived at by natural selection in the landscape of possible genetic capabilities. In this case, the local optimum depends on past history of the species, which inhibits our ability to infer a genome sequence \textit{ab initio} from knowledge of the environment by using thermodynamics or maximum caliber as an inference tool. It is the inability to address all assumptions about appropriate constraints that led Dewar to suggest that, when a maximum entropy model fails, it is usually not the maximum entropy concept that has failed but rather the assumptions used in the model \cite{Dewar2009}. Similar conclusions have been drawn by Agozzino and Dill \cite{Agozzino2019}.

Ultimately, the usefulness of maximum entropy or maximum caliber models for biology will be judged by the insight that they provide by linking physics with its natural laws to biology, and in the predictive power that the concept brings to models or systems. In this regard, an understanding is needed of how a these models are related to non-optimal models, such as one would find in a population. For example, kinetic models using the law of mass action are highly variable depending on the rate parameters that are used. How sensitive are maximum entropy or caliber models to uncertainty in the thermodynamic parameters, such as chemical potentials and equilibrium constants from which rate constants are inferred? What are the ranges of feasible concentrations and rates?

In this report, we derive the formula for entropy production rate for systems of coupled reactions from first principles, and in doing so link the ordinary differential equations describing mass action kinetics with the master equation formulation. 
\textcolor{black}{Throughout the report, we assume that the ground states of the reactants and products are equilibrated with their local environments before and after reactions \cite{hunter_jcp_1993, Jarzynski1997, oono_1998, hatano_2001, trepagnier_pnas_2004} such that the use of reference chemical potentials and Boltzmann probabilities are justified.}
We distinguish entropy production rates from free energy dissipation rates based on whether one is optimizing the reaction free energies or probabilities.  We then demonstrate how to find maximum entropy models using an alternative form of the law of mass action. Having found the maximum entropy solution, we show how to efficiently generate a population of feasible mass action models, and compare the population of mass action models to the maximum entropy model. We show that thermodynamically optimal metabolisms are characterized by having the highest \textit{probability} of high rates of free energy dissipation when compared to similar metabolisms. Importantly, systems having a stationary entropy production rate have a \textit{maximal} change in probability with time when compared to similar systems that differ only by their rate constants. We discuss three sources of variability in models of metabolism: (1) variability of rates, metabolite levels, and free energy dissipation rates that would be found in a general population compared to the maximum entropy model; (2) variability in predictions made by maximum entropy models due to uncertainty in estimated standard free energies of reaction (equivalently, equilibrium constants); and finally, (3) variability in predictions made by maximum entropy models due to the range of ionic strengths that may be found in the cytoplasm of the cell. 

\section{\label{sec:level1}Theory}
\subsection{\label{sec:level2}Entropy Definition}

The entropy as described by Boltzmann is,
\begin{equation}\label{entropy}
    S_B = k_B \log W,
\end{equation}
where $k_B$ is the Boltzmann constant and $W$ stands for \textit{Wahrscheinlichkeit}, or probability \cite{Boltzmann1877, Planck1914}. In a famous example used as a simple demonstration, Boltzmann employed a system in which each independent object had the same probability distribution (independent and identically distributed) and as such the number of permutations $\mathcal{P}$  of the objects could stand in place of the probability $W$. In this case, maximizing the probability is the same as maximizing the number of permutations. However, as Boltzmann also pointed out, when not assuming equal a priori distributions, it is the probability that is maximized as a system evolves, not the permutations. Boltzmann demonstrated this case for the velocity distributions of an ideal gas with different mean velocities along the $x$, $y$ and $z$ axes \cite{Sharp2015}. 

Likewise, for a system of $M$ different chemical species, the probability $\text{Pr} \equiv W$ of observing a state in which each species $i$ is observed to have $n_i$ counts is given by the multinomial probability density,
\begin{equation}\label{probability_general}
    \text{Pr}=\begin{pmatrix}
    N_{tot}
    \\
    n_{1},...,n_{M}
    \end{pmatrix}
    \prod_{i}^{M} \theta_{i}^{n_{i}},
\end{equation}where $N_{tot} = \sum_i n_i$ and the probability $\theta_i$ of each species $i$ is given by the Boltzmann probability,
\begin{equation}\label{probability_i}
    \theta_i = \frac{e^{-\mu_{i}^{\circ} / k_{B}T}}{\sum_{i}^{M} e^{-\mu_{i}^{\circ} / k_{B}T}},
\end{equation}where $\mu_{i}^{\circ}$ is the standard chemical potential and $T$ is the temperature. As mentioned above, $k_{B}$ is the Boltzmann proportionality constant such that $k_{B}T$ is the ambient energy due to the temperature $T.$ For brevity, we will use $\beta = (k_BT)^{-1}$ for the inverse ambient energy. In the case that each species is independent and identically distributed (i.i.d.), the probabilities of all species are equal, i.e. $\theta_i = \theta_j$ for all $i$ and $j$. It follows that the resulting probability density function is again proportional to the number of permutations, as mentioned above. Using a constant of proportionality, $c$, Eqn. \eqref{probability_general} can be simplified to the following,
\begin{align}
    \text{Pr}  & = c {N_{tot}\choose n_{1},\ldots, n_{M}}
    \label{probability_uniform} \\
 & = W \text{(i.i.d)}.
\label{test}
\end{align}
In this case, $W$ is proportional to the number of ways of distributing $N_{tot}$ molecules among $M$ different chemical species. To distinguish between the more general meaning of $W \equiv \text{Pr}$ and the demonstration case of Eqn. \eqref{probability_uniform}, we will use the symbol $S_U$ to specify the case,
\begin{equation}\label{uniform_entropy}
    S_U/k_B = \log W \text{(i.i.d)}.
\end{equation}
It is $S_U$ that is defined as entropy in some textbook descriptions. We are interested in the general multinomial case where $W \equiv \text{Pr}$ such that,
\begin{equation}\label{entropy_prob}
\begin{split}
    S/k_B 
    &= \log \text{Pr} \\
    &= \log \left( \begin{pmatrix}
    N_{tot}
    \\
    n_{1},...,n_{M}
    \end{pmatrix}
    \prod_{i}^{M} \theta_{i}^{n_{i}} \right ).
\end{split}
\end{equation}
It is important to note that thermodynamic entropy obeys an extremum principle such that for any spontaneous process, the entropy increases to the maximum extent possible given the conditions.  $S_U$ only obeys an extremum principle for a system that follows a uniform probability distribution, while $S = k_B\log \Pr$ (e.g., Eqn. \ref{entropy_prob}) obeys an extremum principle generally. Thus, it is $S$ defined by Boltzmann and restated by Planck \cite{Planck1914} that is related to the classical concept of entropy described by Clausius. However, to be clear, $S$ in Eqn \ref{entropy_prob} is not the classical thermodynamic entropy; \textcolor{black}{it is a single configuration \textit{representation} of the entropy of classical thermodynamics - it is not an average value. The average of $S$, as discussed by Gibbs, is the mathematical function that represents the classical thermodynamic entropy in statistical thermodynamics \cite{gibbs_1902}. A common point of confusion is to conflate the definition of entropy from classical thermodynamics, where it is defined by experimental observables, with definitions from statistical thermodynamics. Moreover, in classical thermodynamics it is not the case that entropy requires that the system to be closed or at equilibrium; it is that historically entropy had not been defined experimentally for open, non-equilibrium systems. There is a big difference in meaning between the limitation of a scope of a definition due to simple inability to measure something beyond experimental limits and a \textit{requirement}. However, even this experimental definition has been extended such that entropy in classical thermodynamics can be defined in open, non-equilibrium steady state systems \cite{oono_1998, hatano_2001, trepagnier_pnas_2004}.}

Like $S$ in Eqn \ref{entropy_prob}, a single configuration free energy, $G$, of a given state $\{n_1,...,n_M\}$ can be defined. Specifically, by using Sterling's approximation \textcolor{black}{that $\log (n!) \approx n\log n - n$}, $G$ can be derived from Eqn. \eqref{entropy_prob}. In doing so, it is convenient to represent the normalization factor in Eqn (\ref{probability_i}) as $q_B = \sum_i e^{-\beta\mu_{i}^{\circ}}$: 
\begin{widetext}
\begin{eqnarray}
\log \Pr & = & N_{T}\log N_{T} - N_T -\sum_i n_i \log n_i +\sum_i n_i  +  \sum_i n_i\log(e^{-\beta\mu^{\circ}_i}) - \sum_i n_i\log(q_B) \label{logPr1} \\
 & = &  -\sum_i n_i \log n_i  +  \sum_i n_i\log(e^{-\beta\mu^{\circ}_i}) - \sum_i n_i\log(q_B) + N_{T}\log N_{T}  \nonumber \\
  & = &  -\sum_i n_i \log n_i  +  \sum_i n_i\log(e^{-\beta\mu^{\circ}_i}) - N_T\log\frac{q_B}{N_{T}} \label{logPr2} \\
    & = &  -\sum_i n_i (\log n_i  +  \beta\mu^{\circ}_i) - N_T\log\frac{q_B}{N_{T}} \label{logPr2b}  \\
    & = &  -\beta G - N_T\log\frac{q_B}{N_{T}} \label{logPr3} 
\end{eqnarray}
where, 
\begin{eqnarray}
   G & = & \sum_i n_i (\beta^{-1}\log n_i  +  \mu^{\circ}_i) \nonumber \\
   & = & \sum_i n_i\cdot\mu_i. \label{free_energyC}
\end{eqnarray}
\end{widetext}
Rigorously, $G$ is a free energy density and not the Gibbs free energy, which is the ensemble average value, $\left<G\right>.$

Eqn. \eqref{free_energyC} is simply Eqn. \eqref{logPr3} when the contribution from the total number of particles is ignored.
The term $N_T\log (q_B/N_T)$ in Eqns. \ref{logPr2}-\ref{logPr3} is a constant of integration under steady state conditions. The factor $q_B$ can be thought of as a molecular partition function for a hypothetical particle, a \textit{boltzon}, whose internal energy levels reflect the different energy levels of the individual chemical species \cite{Davidson1962}. The chemical potential for this hypothetical particle is $\beta \mu_B = -\log (q_B/N_T)$.
Then the entropy is proportional by ambient energy $\beta$ to an energy that includes the free energy and a contribution due to differing numbers of total particles,
\begin{equation}\label{entropy_gibbs_energy}
 S=-{\beta G} + {N}_{tot}\cdot \beta \mu_B.
\end{equation}

\subsection{\label{sec:level2}Entropy Production,  Entropy Production Rate and Free Energy Dissipation Rate}
The concept that natural systems optimize their entropy or entropy production rate $dS/dt = d\log Pr/dt $ over time is related to the concept that natural systems maximize the change in probability with respect to time, $dPr/dt$. Over time, natural systems tend to move to the most probable state. Implicit in such statements of maximum entropy is the understanding that there are biological constraints on how fast the system can react and which physical configurations are possible if the system is going to react quickly. For example, the upper limit on enzyme reaction rates is approximately $k_{cat}/K_{M}=10^6s^{-1}$  \cite{Radzicka1995,Schomburg2013}. If a system were to produce chemical species simply by maximizing Eqn. \eqref{entropy_prob}, many chemical species would be in such high concentrations that the cytoplasm of a cell would become glass-like, limiting diffusion and decreasing reaction rates significantly \cite{Britton2020, Cannon2018a}. The idea that enzymes are regulated to ensure they do not over-produce chemical species and thereby adversely impact the solution properties inside the cell was first proposed by Atkinson \cite{Atkinson1969}.

To evaluate either the entropy production or the entropy production rate, $dS/dt$, we need the functional forms of each. Yet the functional form of the time derivative of the entropy, $dS/dt$, is not immediately clear. However, it can be easily derived by considering the infinitesimal entropy produced along a reaction path $\bm{\hat{\xi_j}}$, which measures the extent of reaction $j$, 
\begin{equation}\label{entropy_production_j}
 \frac{dS}{{d\xi_j}} = \frac{\partial \log \text{Pr}}{\partial \xi_j}.
\end{equation}
\textcolor{black}{The extent of the reaction $\xi_j$ determines the the amount $n_i$ of molecule $i$ consumed in reaction $j$ through the stoichiomentric coefficient $\gamma_{i,j}$,
\begin{equation}
    \gamma_{i,j} = \frac{\partial n_i}{\partial \xi_j}.
\end{equation}}
Since in chemical reactions, both forward and reverse reactions happen simultaneously but may differ in proportion, each $\xi_j$ below represents the net extent of the reaction such that $\xi_j = \xi_{+j} - \xi_{-j}$, where the signed index symbols on $\xi_{+j}$ and $\xi_{-j}$ represent the forward and reverse reactions, respectively. 

\textcolor{black}{
Generally, in a system of $Z$ reactions, we have unit vectors along each reaction coordinate, $\bm{\hat{\xi_1}}, ..., \bm{\hat{\xi_Z}}$. We define a unit vector for the system of reactions $\bm{\xi} = (\xi_1\bm{\hat{\xi_1}} + \hdots + \xi_Z\bm{\hat{\xi_Z}})$, such that a distance $x$ that measures the extent of reaction of the system along $\bm{\xi}$ is,
\begin{eqnarray}
    \bm{\xi} x & = & [\xi_1, ..., \xi_Z ]^Tx.
\end{eqnarray}
}
The entropy production of a system of $Z$ coupled reactions along the path $\bm{\xi}$ is then given by the derivative along $\bm{\xi}$,
\begin{equation}\label{entropy_production}
\frac{dS}{dx} = \bm{\xi}^T \cdot \nabla S  = \sum_{\substack{j=1}}^{Z} \xi_j \frac{\partial \log \text{Pr}}{\partial \xi_j},
\end{equation}
where the gradient of the entropy $\nabla S$ is the entropy production vector,
\begin{eqnarray}
    \nabla S & = &
    \begin{bmatrix}
        \frac{\partial \log \text{Pr}}{\partial \xi_1} \\
        \vdots \\
        \frac{\partial \log \text{Pr}}{\partial \xi_Z}
    \end{bmatrix} \\
    & = &
    \begin{bmatrix}
        -\frac{\partial\beta G}{\partial \xi_1} + \sum_{i} \gamma_{i,1} \mu_B \\
        \vdots \\
        -\frac{\partial\beta G}{\partial \xi_Z} + \sum_{i} \gamma_{i,Z} \mu_B \\
    \end{bmatrix}.
    \label{entropy_production_vector}
\end{eqnarray}
\textcolor{black}{Note that in the case that $\sum_i \gamma_{i,j} = 0$ for each reaction $j$, $\nabla S = \nabla G$ which is the vector of reaction affinities,
\begin{eqnarray}
    \nabla G & = &
    \begin{bmatrix}
        \frac{\partial G}{\partial \xi_1} \\
        \vdots \\
        \frac{\partial G}{\partial \xi_Z}
    \end{bmatrix}.
    \label{free_energy_production_vector}
\end{eqnarray}
}

Likewise, the entropy production rate is given by the scalar product of product of the gradient and a unit velocity vector, $\bm{\dot{\xi}} = [\frac{d\xi_1}{dt}\bm{\hat{\xi_1}},...,\frac{d\xi_Z}{dt}\bm{\hat{\xi_Z}}]^T$,
\begin{equation}\label{entropy_probability}
\frac{dS}{dt} = \dot{\bm{\xi}}^T \cdot \nabla S  = \sum_{\substack{j=1}}^{Z} \frac{d\xi_j}{dt} \frac{\partial \log \text{Pr}}{\partial \xi_j}.
\end{equation}
For brevity, below we will use the notation $\frac{d\xi_j}{dt} = \dot{\xi}_j$ for the reaction rates.

Substituting Eqn. \eqref{logPr3} and the fact that $N_{tot}=\sum_i n_i$, differentiation provides the simplified set of equations for the entropy production and the entropy production rate:
\begin{equation}\label{entropy_prob_expanded}
\begin{split}
 \frac{dS}{dx} &=\sum_{\substack{j=1}}^{Z} \xi_j \frac{\partial}{\partial \xi_j}
    \left(
    -\beta G + 
    N_{tot}\beta\mu_B
    \right) \\
    &=\beta \sum_{\substack{j=1}}^{Z} \xi_j 
    \left(\frac{-\partial G}{\partial \xi_j} + \sum_{i} \gamma_{i,j} \mu_B
    \right), \\
\end{split}
\end{equation}
and,
\begin{eqnarray}
        \frac{dS}{dt} &  
    = & \beta \sum_{\substack{j=1}}^{Z}
    \dot{\xi}_j \left(\frac{-\partial G}{\partial \xi_j} + \sum_{i} \gamma_{i,j} \mu_B
    \right).
\end{eqnarray}
Here it is assumed that the change in $N_T$ due to a single reaction is such that $\mu_B$ is essentially constant.
Finally, substituting the identities $\partial \beta G / \partial \xi_j = - \log K_j Q_{j}^{-1}$ \textcolor{black}{\cite{Cannon2017}}, where \textcolor{black}{$K_j$ is the equilibrium constant,} and $Q_j$ is the reaction quotient,  the following equalities are obtained:   	
\begin{eqnarray}
\frac{dS}{dx}  & = & \sum_{\substack{j=1}}^{Z} \xi_j \left( \log K_{j} Q_{j}^{-1} + \sum_{i} \gamma_{i,j} \beta \mu_B \right) \label{entropy_production2}\\
\frac{dS}{dt} & = & \sum_{\substack{j=1}}^{Z} \dot{\xi}_j \left( \log K_{j} Q_{j}^{-1} + \sum_{i} \gamma_{i,j} \beta \mu_B \right) \label{entropy_flux}\\
& = &\beta \left( -\frac{dG}{dt} + \frac{d\mu_B}{dt} \right).
\end{eqnarray}
It is clear that maximizing the entropy production, Eqn \ref{entropy_production2}, will not necessarily lead to the same result as maximizing the entropy production rate, Eqn \ref{entropy_flux}, due to the difference between the extent of a reaction and its time derivative. The entropy production is a measure of the amount of entropy produced per reaction, while the entropy production rate is the entropy produced per time. An important question then is, when seeking to model nature, should a computational model maximize the entropy, the entropy production or the entropy production rate? 

The entropy production rate has two contributing terms \--- the free energy dissipation rate and the rate of work to introduce or remove particles to the system. Rigorously, $dG/d\xi_{j} = \beta^{-1} \log K_{j}Q^{-1}_j$ is the reaction affinity for the $j^{th}$ reaction, but for steady state systems with a large number of particles, $\log K_{j} Q_{j}^{-1} \approx -\beta \Delta G $ is the reaction free energy change. For these systems, Eqn. \eqref{entropy_flux} states that the rate of change of the entropy is proportional to the sum over all reactions of the product of the reaction rate and the negative of the free energy for each reaction. In the case of steady state systems with ensemble averaging and $\sum_i \gamma_{i,j} = 0$, the entropy production rate from Eqn. \eqref{entropy_flux} is the negative of the free energy dissipation rate, $dS/dt = -d\beta G/dt$ \cite{Ge2010}. Furthermore, the entropy production rate is related to the real-valued, continuous chemical master equation since,
\begin{equation}\label{entropy_prob_pr}
\begin{split}
    \frac{d \log \text{Pr}}{dt}
    &=\frac{1}{\text{Pr}} \frac{d \text{Pr}}{dt}.
\end{split}
\end{equation}
Therefore, the change in probability with respect to time due to reaction $j$ is,    
\begin{eqnarray}
    \frac{d \text{Pr}}{dt}
    &=& \text{Pr} \left( 
    \dot{\xi}_j \log K_{j}Q_{j}^{-1} + \dot{\xi}_j \sum_{i} \gamma_{i,j} \beta \mu_B
    \right) \label{dPr_dt} \\
    &=& \text{Pr} 
     \left( 
    \frac{-dG}{dt} + \frac{d\mu_B}{dt}
    \right) \beta.    \label{dPr_dt_terms}
\end{eqnarray}
The time-dependent probability considering all reactions is,
\begin{multline}
\label{master_eqn}
    \text{Pr}(t+ \Delta t)
    = \text{Pr}(t)\sum_{\substack{j=1}}^{Z} \exp 
    \left(\int_{t}^{t +\Delta t}\left[ \dot{\xi}_j \cdot\log K_{j}Q_{j}(t) + \right. \right. \\
     \left. \left. \dot{\xi}_j \cdot\beta  \mu_B \sum_{i} \gamma_{i,j} \right] dt \right),
\end{multline}
which is the chemical master equation. The transition probability for each reaction $j$ is the respective exponential term, which can be shown to be a conditional probability. The arguement of the exponential is the entropy production rate for reaction $j$. For the first term of the integrand, integration from $t = 0$ to the time $\Delta t$ for one stoichiometric reaction at steady state ($\dot{\xi}_j$ constant) gives,
\begin{align}
\label{G(t)}
 \int_{t=0}^{t=\Delta t} \dot{\xi}_j \cdot\log K_{j}Q_{j}^{-1}(\xi_{j}(t)) dt
     = &\text{\ \ \ \ \ \ \ \ \ \ \ \ \ \ \ \ } \\
\text{\ \ \ \ \ \ \ \ }    \dot{\xi}_j \left[\log K_{j} 
      - \sum_{i} \gamma_{i,j} \log (n_i(0) +\gamma_{i,j}) \right]&\Delta t \nonumber \\
     = & -\dot{\xi}_j \Delta t \cdot \beta \Delta G_{j} \text{\ \ \ \ \ \ \ \ \ \ \ \ \ \ \ \ } 
\end{align}
so that the master equation due to the firing of just reaction $j$ is,
\begin{eqnarray}
\label{master_eqn_one_reaction}
    \text{Pr}(\Delta t) &=& \text{Pr}(0)\cdot 
     e^{-\dot{\xi}_j \Delta t \cdot \beta \Delta G_{j}}
    \cdot e^{\dot{\xi}_j \Delta t \cdot \beta \mu_B \sum_{i} \gamma_{i,j}}.
\end{eqnarray}
\textcolor{black}{When evaluating the master equation for a change to an adjacent state such as the firing of only one or several reactions, the error due to using Sterling's approximation is negligible. However, when the change of state is large such as going from a highly improbable state to a steady state, the error may be significant. In this case it is more convenient to simply evaluate the difference in entropies using Eqn \eqref{entropy_prob} and the gamma or log-gamma function.}

\subsection{\label{sec:most_prob_state}The Most Probable State: Maximum Entropy}
The most probable state is found by maximizing the probability, Eqn \eqref{probability_general}, or equivalently the logarithm of the probability, Eqn \eqref{entropy_prob}, subject to the constraint that a finite number of reactions may occur \--- that is, by applying Lagrange's method of undetermined multipliers to constrain the extent of each reaction to $\xi_j = c_j$, a constant, 
\begin{eqnarray}
F & = & \log \text{Pr} - \lambda \left(\sum_j \xi_j -c_j\right)
\label{lagrange_form1}
\end{eqnarray}
where $\lambda$ is the undetermined multiplier \textcolor{black}{and $\log \text{Pr} = S$ is the Boltzmann entropy from Eqn \ref{entropy_prob}.
Taking the derivative of $F$ with respect to the extent of reaction of the system $x$, 
\begin{widetext}
    \begin{eqnarray}
   \frac{dF}{dx} & = & \begin{bmatrix}
 \xi_1 (\log \left[ K_1Q_1^{-1}\left(e^{\beta\mu_B}\right)^{\sum_i \gamma_{i,1}}\right] - \lambda)\\
\vdots & \\
 \xi_Z (\log \left[ K_ZQ_Z^{-1}\left(e^{\beta\mu_B}\right)^{\sum_i \gamma_{i,Z}}\right] - \lambda)
\end{bmatrix}^T  \cdot \bm{1}
=  \begin{bmatrix} 0 \\ \vdots \\ 0 \end{bmatrix}^T \cdot \bm{1} \label{max_entropy_production}
\end{eqnarray}
\end{widetext}
where again the $\xi_j$ are the values for the extent of reaction in the unit vector $\bm{\xi}$ and $\bm{\lambda} = [\lambda, ..., \lambda]^T$ is the vector containing the constraint variable. The constraint
in terms of the gradient of the entropy is,
\begin{eqnarray}
   \nabla \bm{S} & = & \bm{\lambda}.
\end{eqnarray}
}
\textcolor{black}{If it is the case that the sum of the stoichiometric coefficients for each reaction $j$ is zero, $\sum_i \gamma_{i,j} = 0$, then it can be seen that the constraint $\lambda$ corresponds to the reaction affinity,
\begin{eqnarray}
   \nabla \bm{G} & = & \bm{\lambda}.
\end{eqnarray}
The specific value of $\lambda$ will of course depend on the system conditions, especially the boundary conditions that determine whether the system is at equilibrium or non-equilibrium.}

Thus, the state with maximum probability has the property that for each reaction $j$,
\textcolor{black}{
\begin{equation}
{\xi}_j\left(\log \left[K_jQ_j^{-1}\left(e^{\beta\mu_B}\right)^{\sum_i \gamma_{i,j}}\right] - \lambda\right) = 0,
\label{eq:most_probable_criteria}
\end{equation}
}
such that either a reaction $j$ will have, 
\begin{equation}
{\xi}_j = 0, \label{constraint2_max}
\end{equation}
or,
\begin{equation}
\log \left[ K_jQ_j^{-1}\left(e^{\beta\mu_B}\right)^{\sum_i \gamma_{i,j}}\right] = \lambda. \label{constraint1_max}
\end{equation}
That is, either a reaction is at equilibrium such that Eqn \ref{constraint2_max} is true, or reactions $j$ not at equilibrium all have the same change in entropy with respect to the extent of reaction.
Analogous to Boltzmann's original formulation of entropy in which each energy microstate of independent particles is assumed to be equally likely, each reaction is assumed to be equally likely from an energy perspective. In other words, the configuration that has the highest density is the state in which all reactions occur with the same probability. 

Likewise, the state of time-stationary probability has the property that the time derivative of $F$ is zero such that,
\begin{widetext}
\begin{equation}
 \frac{dF}{dt}  = 
\begin{bmatrix}
 \dot{\xi}_1 (\log \left[ K_1Q_1^{-1}\left(e^{\beta\mu_B}\right)^{\sum_i \gamma_{i,1}}\right] - \lambda)\\
\vdots & \\
\dot{\xi}_Z (\log \left[ K_ZQ_Z^{-1}\left(e^{\beta\mu_B}\right)^{\sum_i \gamma_{i,Z}}\right] - \lambda)
\end{bmatrix}^T \cdot \bm{1}  =  \begin{bmatrix} 0 \\ \vdots \\ 0 \end{bmatrix}^T \cdot \bm{1}
\label{lagrange_solution2}
\end{equation}
\end{widetext}
The requirements for the most stable state are that either Eqn \eqref{constraint1_max} is obeyed as before or,
\begin{equation}
\dot{\xi}_j = 0 \label{constraint3_max}.
\end{equation}
For a steady state, there is an additional requirement that the species counts $n_i$ (or concentrations) be stable with respect to time such that $\mathbf{S}\cdot\bm{\dot{\xi}}^{ss} = 0$ where $\mathbf{S}$ is the stoichiometric matrix and $\bm{\dot{\xi}}^{ss}$ is the vector of steady state reaction fluxes. $\bm{\dot{\xi}}^{ss}$ may differ from the unit reaction fluxes $\bm{\dot{\xi}}$ due to balancing of reaction stoichiometries to obtain mass conservation in the steady state.  Consequently, according to Eqn \eqref{lagrange_solution2}, the maximum entropy state is the state in which any reaction $j$ has the entropy production value $\lambda$, otherwise $\dot{\xi}_j = 0$ such that reaction $j$ is at equilibrium.

\textcolor{black}{
One must keep in mind that the analysis above indicates which state is the most probable; this is the state of maximum entropy. The analysis
does not indicate which state is the state having the maximum entropy production or the maximum entropy production rate. There are two conditions that must be kept in mind: 
\begin{itemize}
\item \textit{Condition 1}: If the boundary conditions are the same for every state such that the total drop in entropy is $Z\lambda$ and if all states have the same steady state rate, then all states have the same entropy production rate $\frac{dS}{dt}$. 
\item \textit{Condition 2}: If the boundary conditions are the same for every state such that the total drop in entropy is $Z\lambda$ but the states have different steady state rates, it is not necessarily the case that the state obeying Eqns \ref{constraint2_max} and \ref{constraint1_max} will have the maximum entropy production rate $\frac{dS}{dt}$.
\end{itemize}}

\textcolor{black}{
To answer the question asked above, whether a computational model maximize the entropy, the entropy production or the entropy production rate, it is the entropy that should be maximized. 
While this may seem counter-intuitive -- that the conditions above do not result in a state of maximum entropy production or maximum entropy production rate, there is a subtle semantic issue at hand. Specifically, 
if instead of defining probability densities based on species counts $n_i$, 
one were to instead define density functions $f$ based on reactions $\xi_j$ such that,
\begin{equation}
    f({\xi}_j) = \xi_j \log \left[ K_jQ_j^{-1}\left(e^{\beta\mu_B}\right)^{\sum_i \gamma_{i,j}}\right]
\end{equation}
or,
\begin{equation}
    f(\dot{\xi}_j) = \dot{\xi}_j \log \left[ K_jQ_j^{-1}\left(e^{\beta\mu_B}\right)^{\sum_i \gamma_{i,j}}\right]
\end{equation}
and instead of using the Boltzmann entropy of Eqn \ref{entropy_prob} used Boltzmann's H-theorem as a definition of entropy, for instance,
\begin{equation}
     -H(\dot{\xi}) = -\sum_{j=1}^Z f(\dot{\xi}_{j}) \log f(\dot{\xi}_{j}), \label{eq:h-theorem1}
\end{equation}
then the maximum entropy production and entropy production rate states by these definitions would indeed be the most probable states. Note that Eqn \eqref{eq:h-theorem1} is a thermodynamic function; Boltzmann's original H-theorem is not necessarily as the densities $f$ were observational rather than thermodynamic in nature, making the original H-theorem more closely related to (but distinct from) the Maximum Caliber concept \cite{Jaynes1985, Presse2013}. The observational densities $f$ used by Boltzmann would presumably have the same values as $f(\dot{\xi}_j)$ above when the observational densities are sampled over the ergodic time scale.
See section \ref{sec:empirical_prob} for further discussion.
}

The maximum entropy state is the state at which the probability density is maximized; it can be thought of as a reference state. Denoting the vector of reaction quotients $\bm{Q}^{\ast} = \left[Q{\ast}_{1},...,Q{\ast}_{Z}\right]^{T}$ for this reference state and the vector of equilibrium constants $\bm{K} = \left[K_{1},...,K_{Z}\right]^{T}$, the vector of chemical potentials $\bm{\mu}{\ast} = \left[\mu{\ast}_1,...,\mu{\ast}_Z\right]^{T}$ for the reference state are calculated using the stoichiometric matrix $\mathbf{S}$,
\begin{equation}
    \beta \bm{\mu}^{\ast} = \mathbf{S}^{-1}\cdot\log{(\bm{K}^{T}\circ \bm{Q}^{\ast -1})}, \label{eq:reference_potentials}
\end{equation}
\noindent where $\circ$ indicates the Hadamard element-wise product.

\textcolor{black}{
\subsection{\label{sec:level2} Maximum Entropy is the State of Maximum $dPr/dt$}
Generally in any system, whether chemical or not, we may be interested in finding the state that has the greatest change in probability with time, $d Pr/dt$, given by Eqn \eqref{dPr_dt}, but constrained such that the rates $\bm{\dot{\xi}}$ are at steady state. Similar to the procedure above, this state can be found using Lagrange's method of undetermined multipliers in which we define a new function $F'$ involving a steady state constraint such that,
\begin{equation}
    F' = \frac{d\text{Pr}}{dt} - \lambda' (\sum_{j=1}^Z (\dot{\xi_j} -c_j)).
    \label{eq:H}
\end{equation}
where the constraints $\sum_{j=1}^Z (\dot{\xi}_j -c_j)$ are the assurance that each process rate $\dot{\xi}_j$ is constant ($c_j$) for all processes $j$. 
}

\textcolor{black}{
Since,
\begin{eqnarray}
    \frac{d\text{Pr}}{dt} & = & 
    \begin{bmatrix}
    \text{Pr} \frac{\partial \log Pr}{\partial \xi_1}\dot{\xi}_1 \\
    \vdots \\
    \text{Pr} \frac{\partial \log Pr}{\partial \xi_Z}\dot{\xi}_Z \\
    \end{bmatrix}^T \cdot \bm{1}
     =  \begin{bmatrix}
    \text{Pr} \frac{\partial S}{\partial \xi_1}\dot{\xi}_1 \\
    \vdots \\
    \text{Pr} \frac{\partial S}{\partial \xi_Z}\dot{\xi}_Z \\
    \end{bmatrix}^T \cdot \bm{1},
\end{eqnarray}
taking the gradient $\nabla_{\bm{\dot{\xi}}}$ of Eqn \eqref{eq:H} with respect to the reaction fluxes $\dot{\xi}_j$, the state of maximum change in probabillity with time leads to the condition that either $\dot{\xi}_j = 0$ or,
\begin{eqnarray}
    \begin{bmatrix}
    \text{Pr} \frac{\partial S}{\partial \xi_1} - \lambda'\\
    \vdots \\
    \text{Pr} \frac{\partial S}{\partial \xi_Z} - \lambda'\\
    \end{bmatrix} & = &
    \begin{bmatrix}
        0 \\
        \vdots \\
        0
    \end{bmatrix}
    \label{lagrange_solution_dprdt1}
\end{eqnarray}
Since Pr is a constant due to a specific steady state, Eqn \eqref{lagrange_solution_dprdt1} can be rewritten as,
\begin{eqnarray}
    \begin{bmatrix}
    \frac{\partial S}{\partial \xi_j}\\
    \vdots \\
    \frac{\partial S}{\partial \xi_j}\\
    \end{bmatrix} & = &
    \begin{bmatrix}
        \lambda \\
        \vdots \\
        \lambda
    \end{bmatrix}
        \label{lagrange_solution_dprdt2}
\end{eqnarray}
where $\lambda = \lambda'/\text{Pr}$. Here, the constraint once again is simply $\nabla S = \lambda$, and Eqn \eqref{lagrange_solution_dprdt2} is the most probable state condition. Because of \textit{Condition 2} discussed above, it is not the case that the state with the maximum entropy production rate $\frac{dS}{dt}$ due to a high steady state rate also maximizes the change in probability with time. The change in probability with time is maximized under the maximum entropy conditions for those processes $j$ in which $\dot{\xi}_j \neq 0$. Note that Eqns \ref{eq:H}-\ref{lagrange_solution_dprdt2} apply generally, not just to chemical reaction systems.}

\subsection{\label{sec:level2}Mass Action Kinetic Models from Maximum Entropy}
\textcolor{black}{The practical question then is how to develop a maximum entropy production model for a biological system such as metabolism? The maximum entropy condition of Eqn \eqref{lagrange_solution2} can be solved directly using mathematical optimization with constraints. However, mathematical optimization does not provide any additional physical insight into the maximum entropy state. Fortunately, following the development of the Arrehnius equation there was a concerted effort to develop a thermodynamic formulation of mass action. That effort ultimately eventually led to the development of transition state theory \cite{Wynne-Jones1935,Laidler1983}. An important conceptual step towards this goal was Marcelin’s equation for mass action \cite{Marcelin1910}, which, although it was ultimately shown to be incorrect, can be used to find a maximum entropy kinetic model that provides additional physical insight into the maximum entropy requirements. }

Consider a reaction involving $n_A$ molecules of reactants $A$, and $n_B$ molecules of products $B$, each with respective unsigned stoichiometric coefficients $\nu_{i,j} = |\gamma_{i,j}|$ for each molecular species,
\begin{equation}\label{reaction_basic}
\ce{\nu_{A,1}n_{A}  <=>[k_{1}][k_{-1}] \nu_{B,1}n_{B} }.
\end{equation}
The usual kinetic law of mass action for the net reaction flux is,
\begin{equation}\label{kinetic_law}
\dot{\xi}_{1} = k_{1}n_{A}^{\nu_{A,1}}  - k_{-1}n_{B}^{\nu_{B,1}},
\end{equation}
where $k_1$ and $k_{-1}$ are constants of proportionality (rate parameters). The challenge is to find a maximum entropy steady state using the law of mass action, yet the usual rate laws such as Eqn. \eqref{kinetic_law} do not contain any thermodynamic terms to optimize. However, a mass action rate law such as Eqn. \eqref{kinetic_law} can be turned into a thermo-kinetic equation by a simple algebraic rearrangement that factors each term such that, 
\begin{equation}\label{kinetic_law_expanded}
\begin{split}
    \dot{\xi}_{1} 
    &=  k_{-1}n_{B}^{\nu_{B,1}} 
    \left( 
    \frac{k_{1}n_{A}^{\nu_{A,1}}}{k_{-1}n_{B}^{\nu_{B,1}} }
    \right)
    - k_{1}n_{A}^{\nu_{A,1}}  
    \left(
    \frac{k_{-1}n_{B}^{\nu_{B,1}} }{k_{1}n_{A}^{\nu_{A,1}} }
    \right)\\
    &= k_{-1}n_{B}^{\nu_{B,1}} (K_1 Q_{1}^{-1}) - k_{1}n_{A}^{\nu_{A,1}} (K_{-1} Q_{-1}^{-1}).
\end{split}
\end{equation}
The reaction rate equation now has thermodynamic terms, $K_1 Q_{1}^{-1}$ and $K_{-1} Q_{-1}^{-1}$, and kinetic terms, $k_{-1}n_{B}^{\nu_{B,1}}$ and $k_{1}n_{A}^{\nu_{A,1}}$ \textcolor{black}{($K_{-1} Q_{-1}^{-1}$ is simply $(K_{1} Q_{1}^{-1})^{-1}$.)} Equation \eqref{kinetic_law_expanded} is an exact but alternate representation of the law of mass action, \textcolor{black}{which says that the rate is proportional to the instantaneous thermodynamic odds of the forward reaction - multiplied by the average time before the odds change due to an addition of another reactant molecule - minus the thermodynamic odds of the reverse reaction - again multiplied by the average time before addition of another reactant molecule for the reverse reaction.} Setting each kinetic term to a constant \textcolor{black}{$c$} is equivalent to assuming that each reaction occurs on the same time scale \textcolor{black}{such that the rate equation becomes,
\begin{equation}
        \dot{\xi}_{1} = c (K_1 Q_{1}^{-1}) - c (K_{-1} Q_{-1}^{-1}).
\nonumber
\end{equation}
}
Although not generally correct, this formulation has the advantage that it makes the energy surface convex, since the surface of the log of an exponential distribution is convex for each fixed number of particles, $N_{tot}$. The result of the approximation is the Marcelin equation \cite{Marcelin1910} and can be used to easily find the maximum entropy steady state \cite{Cannon2018a}. (The time dependence can be added back in post-hoc such that a general formulation of the law of mass action is once again obtained). \textcolor{black}{Using this assumption with $c = 1$ at steady state,} a set of $Z$ sequential reactions having equal rates,  
\begin{equation}\label{flux}
    \dot{\xi}_{1} = \dot{\xi}_{2} =...=\dot{\xi}_{Z},
\end{equation}
\textcolor{black}{will have rates that are proportional to the thermodynamic driving forces such that}, 
\begin{equation}\label{fluxes}
\begin{split}
    \dot{\xi}_{1} &= K_{1}Q_{1}^{-1} - K_{-1}Q_{-1}^{-1},\\
    \dot{\xi}_{2} &= K_{2}Q_{2}^{-1} - K_{-2}Q_{-2}^{-1},\\
    &... ,\\
    \dot{\xi}_{Z} &= K_{Z}Q_{Z}^{-1} - K_{-Z}Q_{-Z}^{-1}.
\end{split}
\end{equation}
The product of the $M$ by $Z$ stoichiometric matrix, $\mathbf{S}$, and the vector of reaction fluxes $\dot{\bm{\xi}}$ is zero at steady state,
\begin{equation}\label{optimization}
    \mathbf{S} \cdot \dot{\bm{\xi}} = 0.0.
\end{equation}
The convention is used here is that $\mathbf{S}$ has rows corresponding to metabolites and columns corresponding to reactions. \textcolor{black}{Eqn \eqref{optimization} enforces mass balance at steady state, in which case the reaction flux may be a multiple $\alpha$ of the unit reaction flux, $\dot{\xi}_j^{ss} = \alpha \dot{\xi}_j$. In this case, the steady state rates are such that,
\begin{equation}
    \dot{\xi}_j^{ss} = (\alpha K_{j}Q_{j}^{-1} - \alpha^{-1}K_{-j}Q_{-j}^{-1})/\alpha. 
    \label{marcelin_rate_law}
\end{equation}
Consequently, the constraints of the most probable state, Eqns \eqref{constraint1_max} or \eqref{constraint3_max}, may not be strictly enforced since a rate $\dot{\xi}_j^{ss}$ may not be single valued. However, as shown in Figure \ref{fig:marcelin_approximation}, the rates are approximately single valued for the cases $\alpha = 1, 2, 3$. 
\begin{figure}
    \centering
    \includegraphics[width=1.0\linewidth]{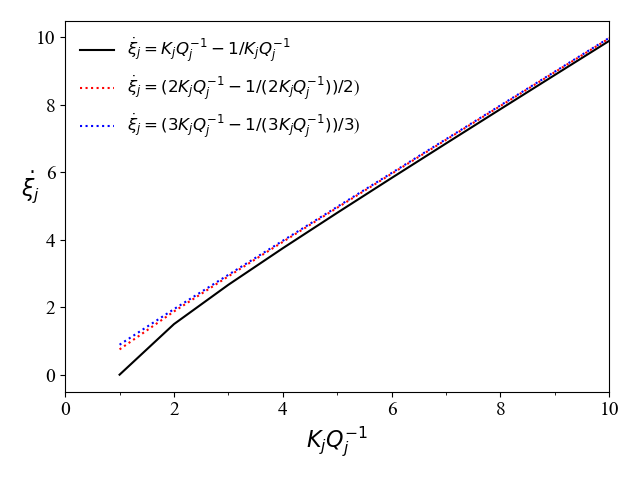}
    \caption{When using the Marcelin approximation (Eqns \eqref{fluxes}) to the mass action rate equations, it is not necessarily the case that all reactions $j$ at steady state will have the same value of the thermodynamic odds $K_jQ_j^{-1}$. The reason for this is that the steady state rate $\dot{\xi}_j^{ss}$ may be as multiple $\alpha$ of the unit rate, $\dot{\xi}_j$. However, even mildly non-equilibrium values of $K_jQ_j^{-1}$ result in approximately the same rate $\dot{\xi}_j^{ss}$, as shown in the plot. For reference, $K_jQ_j^{-1} = 5$ corresponds to a free energy change of $-0.94$ Kcals/M or $-3.99$ KJ/M.} 
    \label{fig:marcelin_approximation}
\end{figure}
}
A steady state solution can be found utilizing the rate laws of Eqn \eqref{fluxes} using a non-linear least squares optimization that results in approximate enforcement of the constraints of Eqns \eqref{constraint1_max} and \eqref{constraint3_max}. The optimization has a runtime of a few seconds in contrast to adaptive time step ODE solvers which can 
take days of CPU time to converge due to the stiffness of the equations as they approach the steady state. 
\textit{The physical insight gained from the use of the Marcelin equation is that in the maximum entropy state, each reaction occurs on the same time scale such that the reaction rates are directly proportional to the thermodynamic odds ($K_JQ_j^{-1}$) of the reaction.}



\textcolor{black}{Regardless of how one obtains the maximum entropy solution,} rate constants for the maximum entropy solution are obtained via the mass action rate laws since the reaction fluxes and steady state concentrations are known. For the chemical reaction similar to that shown in Eqn. \eqref{reaction_basic}, the net flux $\dot{\xi}_{j}$ of reaction $j$ is,
\begin{equation}\label{rate_constant_full}
\begin{split}
    &\dot{\xi}_{j}={{k}_{j}}\prod\limits_{i}^{\text{reactants }j}{{{n}_{i}}} - {{k}_{-j}}\prod\limits_{i}^{\text{reactants -}j}{{{n}_{i}}} \\ 
    &={{k}_{j}}\prod\limits_{i}^{\text{reactants }j}{{{n}_{i}}}\left( 1-\frac{{{k}_{-j}}\prod\limits_{i}^{\text{reactants -}j}{{{n}_{i}}}}{{{k}_{j}}\prod\limits_{i}^{\text{reactants }j}{{{n}_{i}}}} \right) \\ 
    &={{k}_{j}}\prod\limits_{i}^{\text{reactants }j}{{{n}_{i}}}(1-{{K}_{-j}}Q_{-j}^{-1}).  
\end{split}
\end{equation}
Since the fluxes, $\dot{\xi}_{j}$, and metabolite concentrations, $n_i$, are known, the rate constants can be determined, 
\begin{equation}\label{rate_constants1}
\begin{split}
    &{{k}_{j}}= \frac{{\dot{\xi}_{j} }}{{\prod\limits_{i}^{\text{reactants }j}n_i}(1-{{K}_{-j}}Q_{-j}^{-1})}\text{\ \ \ \ \ and} \\ 
    &{{k}_{-j}}= \frac{{{K}_{j}}}{{{k}_{j}}}.  
\end{split}
\end{equation}
The usual mass action ODEs using rate constants, \textit{e.g.}, Eqn \eqref{kinetic_law}, can then be solved using either optimization methods or an ODE solver. The kinetically accessible energy surface is not necessarily convex because of the introduction of the rate constants – each reaction now has its own time dependence. 

\subsection{\label{sec:empirical_prob}Inferring the Thermodynamic Probability of a Kinetic Model}
\textcolor{black}{While the master equation (Eqn \ref{master_eqn}) gives the probability density of a state given the counts or concentrations of molecular species and their rates, determining the probability of a state by solving the master equation involves integrating the change in probability from an initial state (Pr$(t=0)$) to a steady state where $d\text{Pr}(t)/dt$ is constant, which can be challenging.
Thus, it is useful to also have an estimator of the relative probability for evaluating empirical distributions which differ by their steady state rates $\dot{\xi}$. This is the context for which maximum caliber methods were developed \cite{Jaynes1985, Presse2013}. 
In maximum caliber approaches, a probability density of a dynamical property, such as a rate $\dot{\xi_j}$, is often formulated using a function such as,
\begin{eqnarray}
    \log Pr(\dot{\xi}) &  \propto & \sum_j \lambda \dot{\xi}_j -c  
    \label{max_cal_density}
\end{eqnarray}
where $\sum_j \dot{\xi}_j - c$ a constraint on the system, for instance to enforce a steady state condition such that all rates equal a constant $c$, and $\lambda$ is a Lagrange multiplier. The density can be normalized by a cumulative density Q such that,
\begin{eqnarray}
    Pr(\dot{\xi}) & = & \frac{e^{\sum_j \lambda \dot{\xi}_j -c}}{Q}
    \label{max_cal_prob}
\end{eqnarray}
where,
\begin{equation}
    Q = \sum_{s} e^{\sum_j \lambda \dot{\xi}_j(s) -c}.
\end{equation}
Here, the sum is over distinct steady states $s$ and the reaction rate is now shown to be a function of the steady state $s$, $\dot{\xi}_j(s)$.
The result is an empirical density function $Pr(\dot{\xi})$ in which the normalization $Q$ is the cumulative density over the observed steady states $s$. The probability density is simply an empirical density function of observed rates, in which the state having the highest density is also the state that has the average density. Comparing Eqn \eqref{max_cal_density} with Eqn \eqref{lagrange_form1}, the latter contains information on the system energy while the former does not. The probability density $Pr(\dot{\xi})$ of Eqn \eqref{max_cal_prob} is the domain of statistics, not statistical thermodynamics, as it does not follow a thermodynamic extremum principle - it does not maximize the thermodynamic entropy nor minimize the free energy, which Jaynes stated, as well \cite{Jaynes1985}.
}

However, relative thermodynamic probabilities of steady states due to different steady states $s$ and their rates $\dot{\xi}^{ss}(s)$ can be calculated as a function of the distance $d$ from the most probable state, as described by the criteria in Eqn. \ref{eq:most_probable_criteria}. 
Defining a distance $d$ for steady state $s$ from the most probable state as,
\begin{eqnarray}
    d(\dot{\xi}(s)) & = & \sum_j^Z \left[\dot{\xi}_{j}(s)\left(\log K_jQ_j^{-1}(s) - \lambda \right) \right]^2, \\
    & = & \sum_j^Z d_{s,j},
\end{eqnarray}
the probability of a state a distance $d$ away from the most probable state is,
\begin{eqnarray}
    \Pr(\dot{\xi}(s)) & = & \prod_j^Z \frac{1}{\sqrt{2\pi \sigma_{s,j}}}e^{-d_{s,j}/2\sigma_{s,j}} \nonumber\\
         & = & \prod_j^Z f(\dot{\xi}_j(s))  \label{gaussian_distribution}
\end{eqnarray}
The use of the product in this context requires that each of the $Z$ reactions be statistical independent. This is generally the case. While the free energies are related such that the sum of the free energy must be the same as the total free energy, the rate constants are random variables. That is, if the rate parameters $k_{+j}, k_{-j}$ are independently and randomly chosen from a distribution, then since the rate parameter values determine the concentrations and hence the reaction quotients, $Q_j$, the $K_jQ_j^{-1}$ are also random. 

An issue with using Eqn \ref{gaussian_distribution} is that the standard deviation $\sigma_j$ cannot necessarily be reliably estimated.
Instead, the probability densities of the reaction free energies can be inferred non-parametrically using a Gaussian kernel density estimation. 
From a random sample population of $N_{ss}$ steady state solutions in which each steady state solution $s$ has a set of reaction quotients $\{Q_{j=1}(s),...,Q_{j=Z}(s)\}$, the estimated probability $\hat{f}(\dot{\xi}_j(s))$ of observing a difference $d_{s,j} = \log K_jQ^{-1}_{j}(s) - \lambda$ is,
\begin{equation}
    \hat{f}(\dot{\xi}_j(s))=\frac{1}{N_{success}}\sum\limits_{k=1}^{N_{ss}}{\frac{1}{{{h}_{j}}{{\pi }^{1/2}}}{{e}^{{{\left( \frac{{{d}_{s,j}}- {d_{k,j}}}{{{h}_{j}}} \right)}^{2}}}}},
\label{eq:empirical_rxn_pdf_partial}
\end{equation}
where $h_j$ is the kernel scaling parameter or bandwidth.
The estimated relative probability of the steady state $s$ can then expressed as the product of the probabilities for each reaction,
\begin{equation}\label{eq:empirical_rxn_pdf}
    \hat{\Pr} (\dot{\xi}(s))=\prod\limits_{j=1}^{Z}{\hat{f}(\dot{\xi}_j(s))}.
\end{equation}
\textcolor{black}{Here again, it is significant to note that if one considers entropy production to be defined with respect to the densities $f(\dot{\xi}(s))$ for reactions $\xi_j$ using Boltzmann's statistical version of the H-theorem \cite{Boltzmann1877, boltzmann1895, boltzmann1896} as shown in Eqn \eqref{eq:h-theorem1}, then,
\begin{equation}
     -\hat{H}(\dot{\xi}(i)) = -\sum_{j=1}^Z \hat{f}(\dot{\xi}(s)) \log \hat{f}(\dot{\xi}(s)), \label{eq:h-theorem2}
\end{equation}
is an estimate of the thermodynamic $H$ function, Eqn \eqref{eq:h-theorem1}. By this definition the maximum entropy production rate state is the state that matches the conditions for the maximum entropy state discussed below Eqn \eqref{lagrange_solution2}. A similar H-theorem estimate can be analogously defined as the maximum entropy production state in which the $d_{s,j}$ measures the distance in entropy production away from the most probable reaction.}



\section{\label{sec:level1}Application and Evaluation}
\begin{figure}
    \centering
    \includegraphics[width=0.85\linewidth]{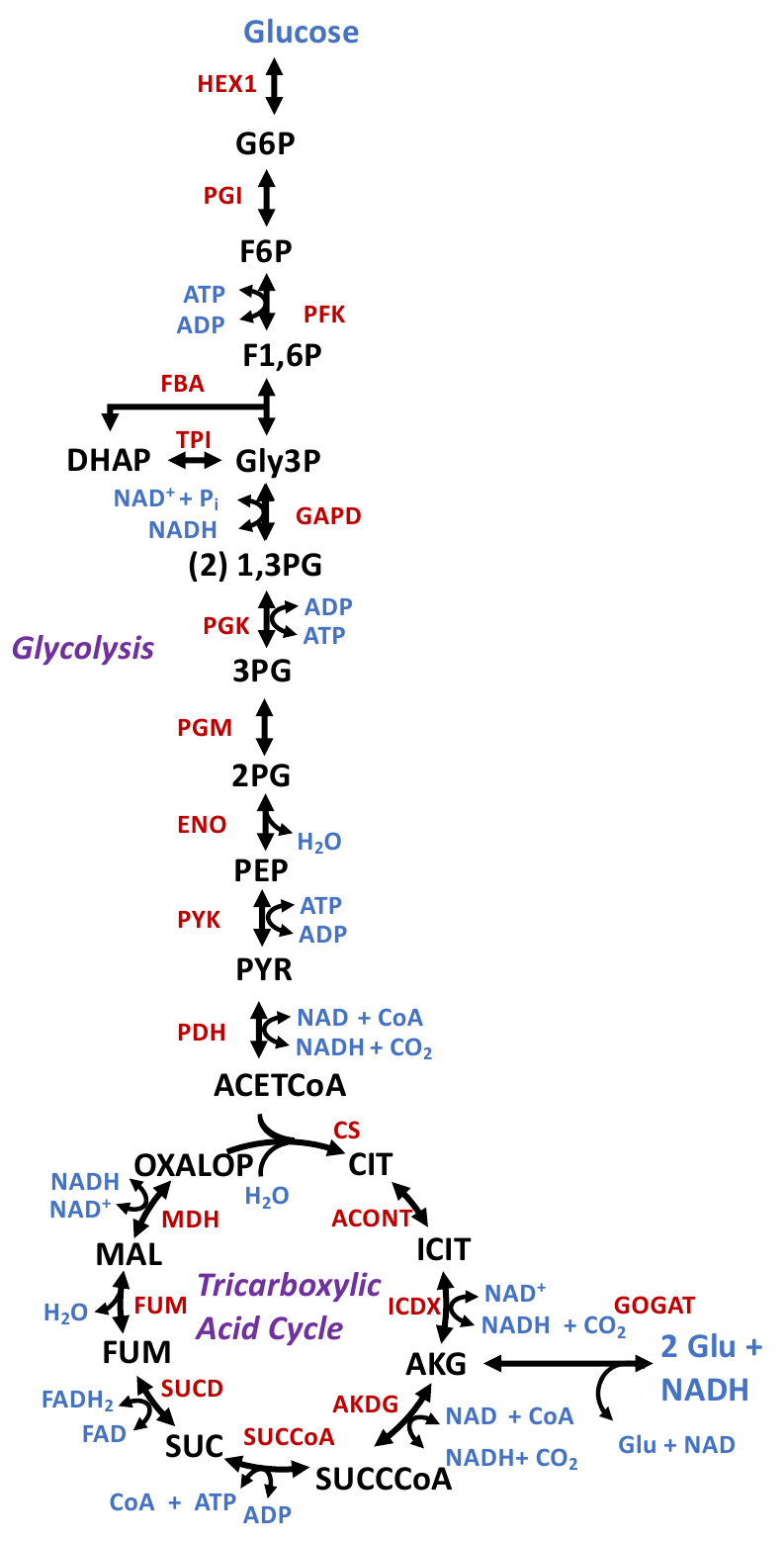}
    \caption{Network of the 21 reactions modeled (not shown:PYRt2m, pyruvate mitochondrial transport reaction).  Metabolites in blue were held fixed during the optimizations and simulations while those in black were variable. The respective enzymes are shown in red. The standard free energies of reaction are provided in the Appendix in Table \ref{tab:variation_Flux_FreeEnergy} (see footnote). \textbf{Metabolites:} G6P: glucose 6-phosphate; F6P: fructose 6-phosphate; F1,6P: fructose 1,6-bisphosphate; DHAP: dihydroxyacetone phosphate; Gly3P: glyceraldehyde 3-phosphate; 1,3PG: 1,3-bisphosphoglycerate; 3PG: 3-phosphoglycerate; 2PG: 2-phosphoglycerate; PEP: phosphoenolpyruvate; PYR: pyruvate; ACETCoA: acetyl-coenzyme A; CIT: citrate: ICIT: isocitrate: AKG: 2-oxoglutarate; SUCCoA: succinyl-coenzyme A; SUC: succinate; FUM: fumarate; MAL: malate: OXALOP: Oxaloacetic acid.     
    \textbf{Enzymes:} HEX1: glucose kinase 1; PGI: phosphoglucoisomerase; PFK: phosphofructokinase; FBA: fructose-bisphosphate aldolase; TPI: triose phosphate isomerase; GAPD: glyceraldehyde phosphate dehydrogenase: PGK: phosphoglycerate kinase; PGM: phosphoglycerate mutase; ENO: enolase; PYK: pyruvate kinase; PDH: pyruvate dehydrogenase; CS: citrate synthase; ACONT: aconitase; ICDX: isocitrate dehydrogenase; GOGAT: glutamine oxoglutarate aminotransferase; AKDG: alpha-ketoglutarate dehydrogenase; SUCCoA: succinyl coenzyme A synthetase; SUCD: succinate dehydrogenase; FUM: fumarase; MDH: malate dehydrogenase.}
    \label{fig:pathway}
\end{figure}

To demonstrate the concepts below, we use a canonical model of central metabolism consisting of glycolysis coupled to the tricarboxylic acid (TCA) cycle shown in Figure \ref{fig:pathway}. In the model, glucose is fed into the system through the glucose kinase reaction (\textit{abbr.} HEX1) and carbon leaves the system as CO$_{2}$ in the TCA cycle. 
Specifically, the metabolic model consisted of the 21 reactions of glycolysis, the tricarboxylic acid (TCA) cycle and the glutamine oxoglutarate aminotransferase reaction (GOGAT) reaction, which includes 37 metabolites, of which 17 were fixed boundary species and 20 were allowed to vary. 
Without the GOGAT reaction, the submatrix of the stoichiometric matrix that describes the TCA cycle is singular (eight reactions but nine variable intermediate metabolites including the input species acetyl CoA). The system is made non-singular by coupling the intermediate $\alpha$-ketoglutarate to a bath using the GOGAT reaction, 2 glutamate + NAD $\Leftrightarrow$ glutamine + $\alpha$-ketoglutarate + NADH, in which glutamate, glutamine, NAD and NADH are held fixed. 

Standard free energies of formation of metabolites in aqueous solution as well as equilibrium constants were determined using the eQuilibrator software \cite{Flamholz2012} using a range of pH and ionic strength values. Rate parameters were determined as described above, Eqn \eqref{rate_constants1}. \textcolor{black}{ Figures and plots below use molar units for free energy such that free energies/mole are used rather than free energies/molecule. The former differs from the latter by Avagadro's number $N_A$. Consequently, Rydberg's constant $R$ multiplied by the temperature $T$, $RT$, is used instead of $\beta$ where  $RT/N_A = \beta^{-1}$). }

In comparison to the maximum entropy model, we focus on three sources of variability in predictions: (1) variability of predictions due to variability in rate constants; (2) variability in predictions due to uncertainty in estimated equilibrium constants; and (3) variability in predictions in models due to the range of ionic strengths that may be found in the cytoplasm of the cell. However, first we describe how variability in steady states arises in biological systems due to variability in nominal rate parameters.
\subsection{\label{sec:level2} Variable Rate Parameters and Implicit Representation of Enzymes.} 
\begin{figure}[hbt!]
\centering
\includegraphics[width=0.5\linewidth]{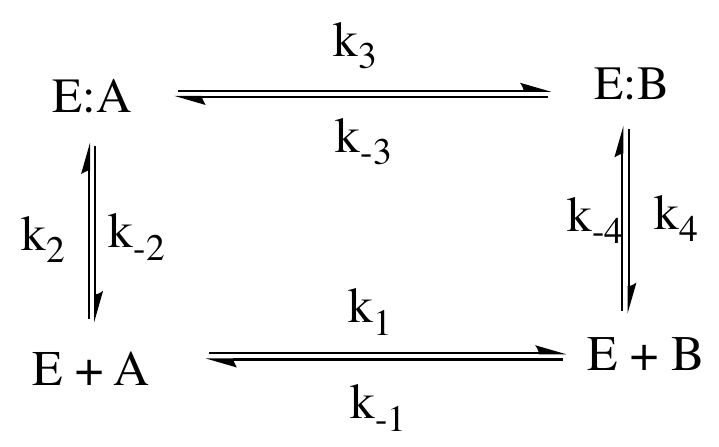}
\caption{Reaction scheme for catalyzed reactions relative to uncatalyzed reactions used in Eqn. \eqref{reaction_basic} }
\label{fig:enzyme_scheme}
\end{figure}
\textcolor{black}{While varying parameters such as mass action rate constants to maximize the entropy of non-equilibrium steady states may not be a familiar concept, this is what biological systems do when under selective pressure (natural selection), even if it is not the entire aspect of natural selection. Through mutation, a set of enzymes in a reaction pathway can be selected for the most thermodynamically efficient enzymes, which ultimately means that a greater amount of energy can be harvested from the environment and put to use to create biomass. }

For enzyme catalyzed reactions such as shown in Figure \ref{fig:enzyme_scheme}, the rate constants determined through Eqn. \ref{rate_constants1} are composite rate constants that implicitly represents an enzymatic process such as the one shown in Figure \ref{fig:enzyme_scheme}. It is not immediately clear how the composite rate constants, $k_{1}$ and $k_{-1}$ in Eqn. \eqref{reaction_basic}, can be related to the elementary enzymatic processes shown in Figure \ref{fig:enzyme_scheme}. The two types of rate constants, composite and elementary, are related through Kirchhoff's voltage law applied to chemical systems \cite{Qian2003},
\begin{equation}
\frac{k_{2}}{k_{-2}} \cdot \frac{k_{3}}{k_{-3}} \cdot \frac{k_{4}}{k_{-4}} = \frac{k_{1}}{k_{-1}},
\end{equation}
such that,
\begin{equation}
k_{-1}\frac{k_{2}}{k_{-2}} \cdot \frac{k_{3}}{k_{-3}} \cdot \frac{k_{4}}{k_{-4}} = k_{1}.
\end{equation}
Likewise,
\begin{equation}
    \label{product_KQ}
    K_{1}Q_{1}^{-1} = K_{2}Q_{2}^{-1} K_{3}Q_{3}^{-1} K_{4}Q_{4}^{-1}.
\end{equation}
As a consequence of generalized detailed balance, the reaction flux $\dot{\xi}_{1}$ represents a coarse graining of the enzyme reaction fluxes according to product of the ratio of the enzyme reaction fluxes,
\begin{equation}
    \dot{\xi}_{+1} = \frac{\dot{\xi}_{+2}}{\dot{\xi}_{-2}} \cdot \frac{\dot{\xi}_{+3}}{\dot{\xi}_{-3} } \cdot \frac{\dot{\xi}_{+4}}{\dot{\xi}_{-4}} \cdot \dot{\xi}_{-1}.
\end{equation}
Thus, while the enzymes are not explicitly represented, the thermodynamics and coarse-grained kinetics of the overall catalyzed process are represented, as expected due to the Haldane relationship between the reaction thermodynamics and enzyme catalysis \cite{Fersht1999}.

\subsection{\label{sec:level2}Comparison of the Maximum Entropy Model to an Ensemble of Mass Action Models}
While the maximum entropy solution is an optimal (\textit{i.e.}, most likely) model, it is not clear how much leeway nature has when selecting organisms with metabolisms that vary from the maximum entropy solution. To investigate this, a population of models with differing reaction rate constants was generated and compared to the maximum entropy solution model. Rate constants were generated by changing metabolite concentrations and thereby altering the value of the reaction quotient $Q$ and solving Eqn \eqref{rate_constants1} for rate constants. For example, for the chemical reaction equation of Eqn \eqref{reaction_basic}, solving the rate law (Eqn \eqref{kinetic_law}) for the rate parameters gives,
\begin{equation}\label{rate_constants2}
\begin{split}
    &{{k}_{1}}=\frac{{\dot{\xi}_{1} }}{{{n}_{A}}(1-{{K}_{-1}}Q_{-1}^{-1})}\text{\ \ \ \ \ and} \\ 
    &{{k}_{-1}}=\frac{{{K}_{1}}}{{{k}_{1}}}.  
\end{split}
\end{equation}
Six ranges of concentrations were used to generate rate constants, corresponding to variation of the concentrations across $r$ orders of magnitude ($10^{\pm r/2}$), centered on the maximum entropy solution, where $r = 0.5, 1, 2, 3, 4$ and $5$. This leads to approximately the same range of variability in rate constants.
\begin{figure*}[t]
\centering
\includegraphics[width=1.0\linewidth]{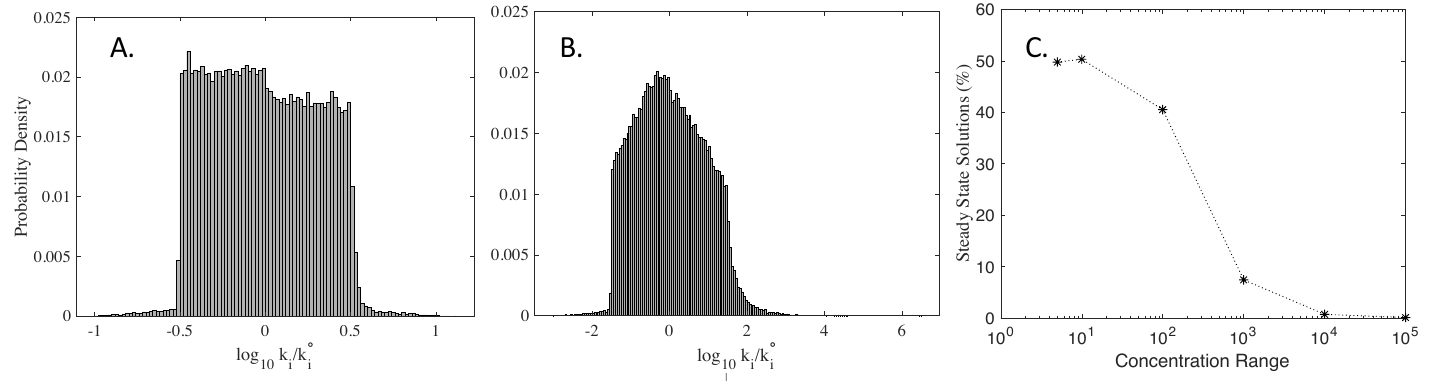}
\caption{(A and B) Probability density function of the log ratio of the population rate constants k$_i$ relative to the respective rate constants k$_i^\circ$ from the maximum  entropy model for each reaction $i$. (A) Rate constants were obtained by varying concentrations from the maximum entropy solution by $10^{\pm 0.5}$-fold. (B) Rate constants were obtained by varying concentrations from the maximum entropy solution by $10^{\pm 1.5}$-fold. (C) Percentage of trials that resulted in a steady state solution as a function of the range over which concentrations were randomly varied. }
\label{fig:histogram}
\end{figure*}

For example, for concentrations varied up or down randomly using $r = 1$ and $3$, the distribution of the ratio of feasible rate constants are shown in Figure \ref{fig:histogram}, A and B, respectively. Each distribution represents the sampling of rate constants over all reactions, except as noted below. Both distributions are asymmetric around the maximum  solution. However, the distribution with the lower variation of range $r = 1$ (Figure \ref{fig:histogram}A) is roughly uniformly distributed on each side of the maximum  solution, while the distribution with higher variation of  $r = 3$ (Figure \ref{fig:histogram}B) falls off steeply on each side of the maximum entropy production solution. 
Since the rate constants for the input boundary reactions (Hex1 and GOGAT) only vary due to variation in the reaction product, not the reactants, they are highly biased to maintain proximity to the rate constants of the maximum entropy solution. These rate constants were therefore removed from both distributions shown in A and B. In contrast, the output boundary reactions producing $\text{CO}_{2}$ or Co-A (PDH, CSM, ICDH, AKDG, SUCCoA) have both reactants and at least one product that varies.

The percent of trials that produced a steady state solution are shown in Figure \ref{fig:histogram}C. Here, results are shown from all six ranges of concentrations used to generate rate constants. As can be seen, the number of feasible solutions decreases rapidly,  from approximately 50\% to less than 10\%, as the allowable concentration range is expanded beyond 10-fold. When the concentration range is  $10^{5}$, only 0.07\% of the random concentrations result in a feasible steady state.

\begin{figure*}[hp]
\centering
\includegraphics[width=0.7\linewidth]{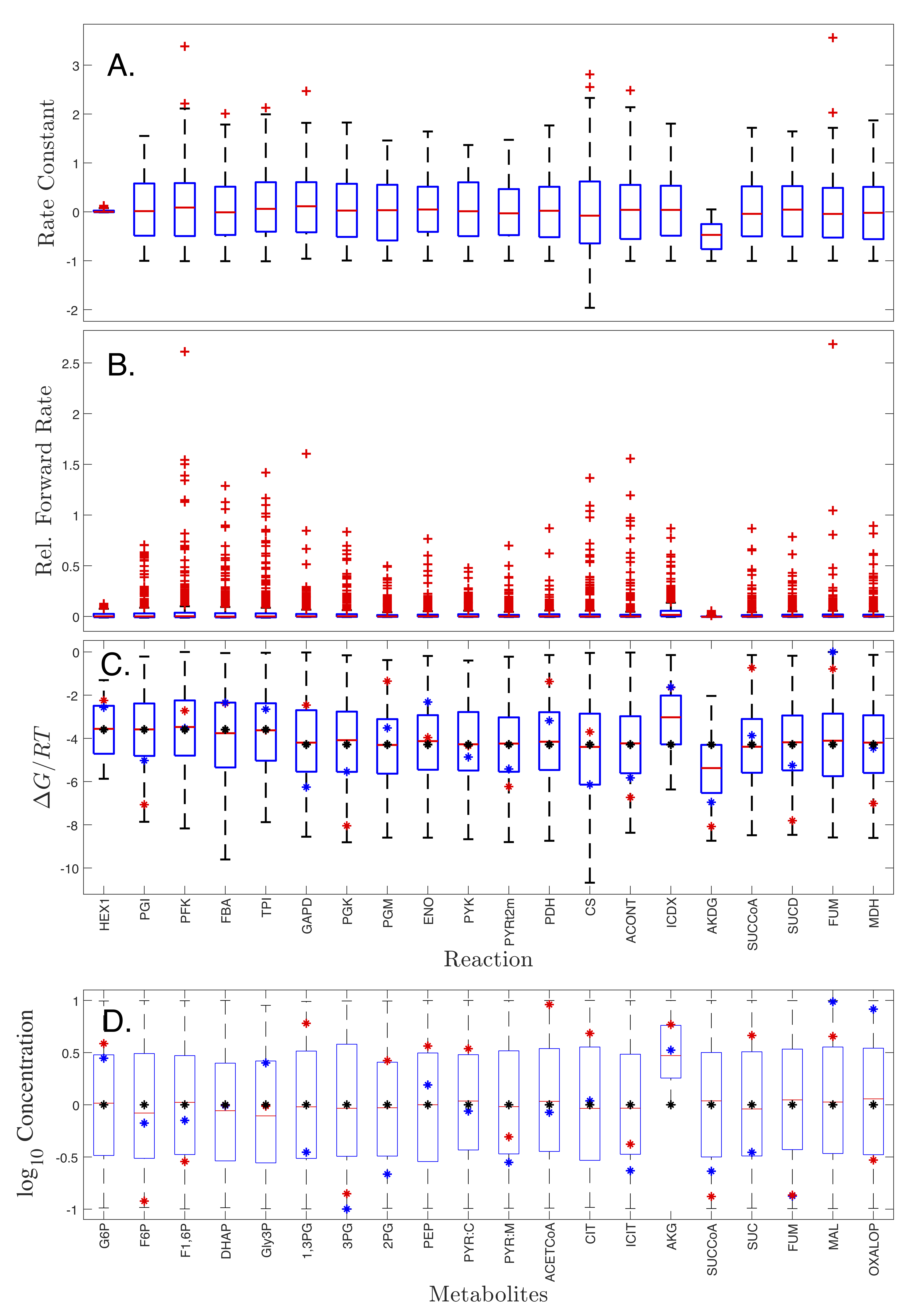}
\caption{Variation in rate constants (A), forward reaction rates (B), and free energy in molar units (C) for each reaction (x-axis) due to alterations in metabolite concentrations (D). Box and whisker plots for each respective reaction indicate average (red line) with $25^{th}$ and $75^{th}$ quantiles (blue boxes). Data points outside a 1.5-fold range covered by the $25^{th}$ to $75^{th}$ quantiles are shown as red '+'s. \textcolor{black}{In (C) and (D) specific models are shown as colored $\ast$'s: the maximum entropy solution  (black $\ast$) is contrasted with the kinetic model which had the largest variance in free energy (red $\ast$) from the maximum entropy solution, and with the solution which had the largest variance in the forward reaction rate (blue $\ast$) from the maximum entropy solution.} \textbf{Abbreviations:} HEX1, Hexokinase; PGI, phosphoglucose isomerase; PFK, phosphofructokinase; FBA, Fructose bisphosphatase; TPI, Triosephosphate isomerase; GAPD, Glyceraldehyde 3-phosphate dehydrogenase; PGK, Phosphoglycerate kinase; PGM, phosphoglycerate mutase; ENO, Enolase; PYK, Pyruvate kinase; PYRt2m, pyruvate transporter; PDH, Pyruvate dehydrogenase; CSM, Citrate Synthase; ACONT, Aconitase; ICDH, Isocitrate dehydrogenase; AKDG, a-ketoglutarate dehydrogenase; SUCOAS, Succinyl-CoA synthetase; SUCD, Succinate dehydrogenase; FUM, Fumarase; MDH, Malate dehydrogenase.}
\label{fig:variation_rates}
\end{figure*}

Properties for individual reactions were evaluated, as well. Figure \ref{fig:variation_rates}A-C shows the resulting distribution of rate constants, reaction rates, and reaction free energies of the ensemble of models generated using sets of random concentrations that varied by up to $10$-fold up or down of the maximum entropy solution ($r=2$: up to 100-fold range across individual concentrations, Figure \ref{fig:variation_rates}D. See also Appendix \ref{app_concs}). Boxplots were used to characterize the resulting rate constants, rates, reaction free energies and concentrations. The ends of each box represents the $25^{th}$ and $75^{th}$ quantiles, denoted $q_n(25)$ and $q_n(75)$, while the whiskers extending from each box denote the most extreme data points falling withing 1.5 times the range between $q_n(25)$ and $q_n(75)$. Data points outside this range are denoted as outliers and are shown as red '+'s.

The feasible rate constants for each reaction are shown in Figure \ref{fig:variation_rates}A as the logarithm of the ratio of the rate constant to the maximum  rate constant. The random rate constants vary asymmetrically around the maximum  values, especially for the glucose kinase (HEX1) reaction, which is an input boundary reaction. As mentioned above, the distribution of rates constants for the input boundary reactions HEX1 and GOGAT (not shown) are considerably skewed due to the fact that only the product concentrations of the reactions are varied.

The analogous reaction rates (for the forward reactions) are shown in Figure \ref{fig:variation_rates}B as the logarithm of the ratio of the rate to the maximum entropy rate. Interestingly, the lowest forward reaction rates across reactions (values of zero) correspond to those for the maximum entropy solution. Reaction free energies varied considerably for each reaction as well, and are shown in Figure \ref{fig:variation_rates}C. The free energies of the maximum entropy solution are shown as black $\ast$, while red $\ast$ marks the steady state in which the reaction free energies have the largest variance away from the maximum entropy solution. Blue $\ast$ denote the distribution that has the largest variance away from the maximum entropy solution with respect to forward reaction rates. As expected, the maximum entropy model due to the Marcelin approximation has the most uniform set of reaction free energies. 

Despite the variation in reaction free energies across the models, each model has the same steady state rate since the rate used to calculate the rate constants (Eqn \eqref{rate_constants1}) was steady state rate from the  the maximum entropy model. Hence, each model extracts the same amount of energy per glucose molecule. The energy extracted per mole of glucose consumed can be calculated by dividing the free energy dissipation rate by the net reaction flux of the glucose uptake reaction, which is the glucose kinase reaction HEX1 in the model,
\begin{align}
 \beta \Delta {{G}_{glucose}} &=\frac{1}{{\dot{\xi}_{\text{glucose uptake}}}}\frac{d\beta G}{dt} \label{delta_G_glucose}\\ 
 & =\frac{-1}{{\dot{\xi}_{Hex}}}\sum_{j=1}^{Z}{{\dot{\xi}_{j}} \log {{K}_{j}}Q_{j}^{-1}}.
 \label{delta_G_glucose2}
\end{align}
The free energy extracted per glucose is a constant, namely $-146.6 \cdot RT$ (-363.4 kJ/mol). Likewise, because each model in the ensemble has the same steady state rate of 36.4 (unitless time) and the same overall free energy change, the free energy dissipation rate also has the same value for each model, -5334 ($\Delta G/RT \cdot$ unitless time), demonstrating Condition 1 in Section \ref{sec:most_prob_state}.

\begin{figure}[hbt]
\centering
\includegraphics[width=0.49\linewidth]{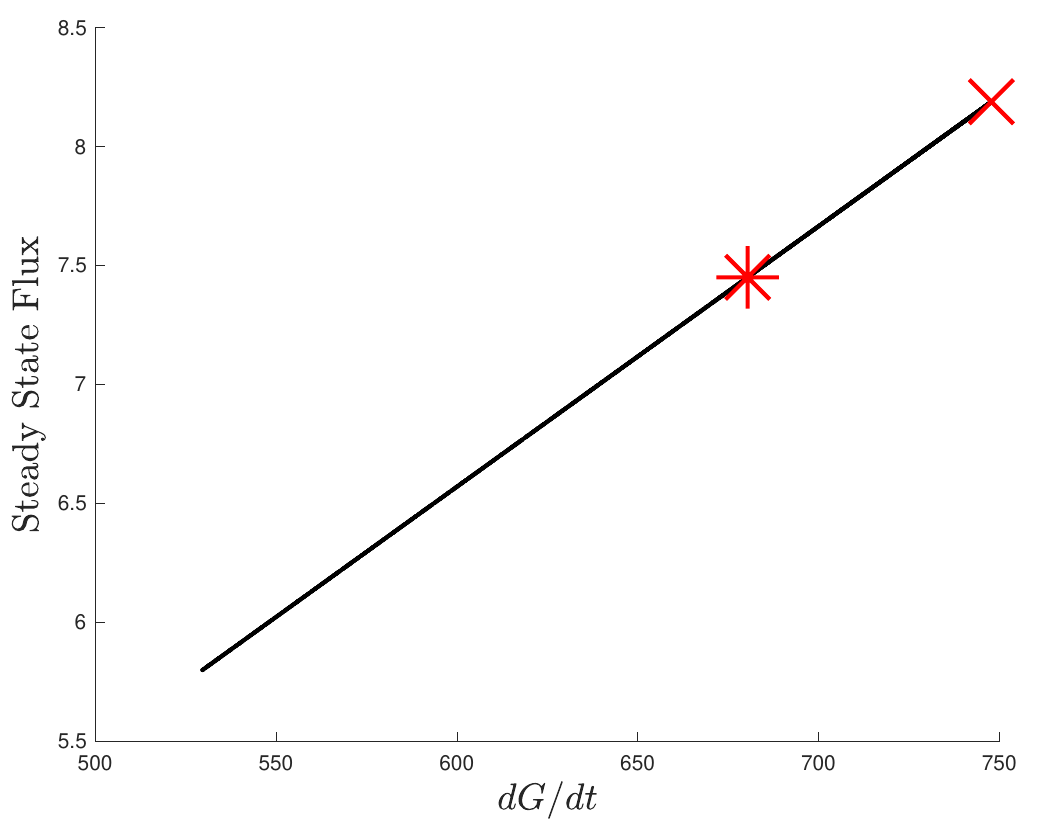}
\includegraphics[width=0.49\linewidth]{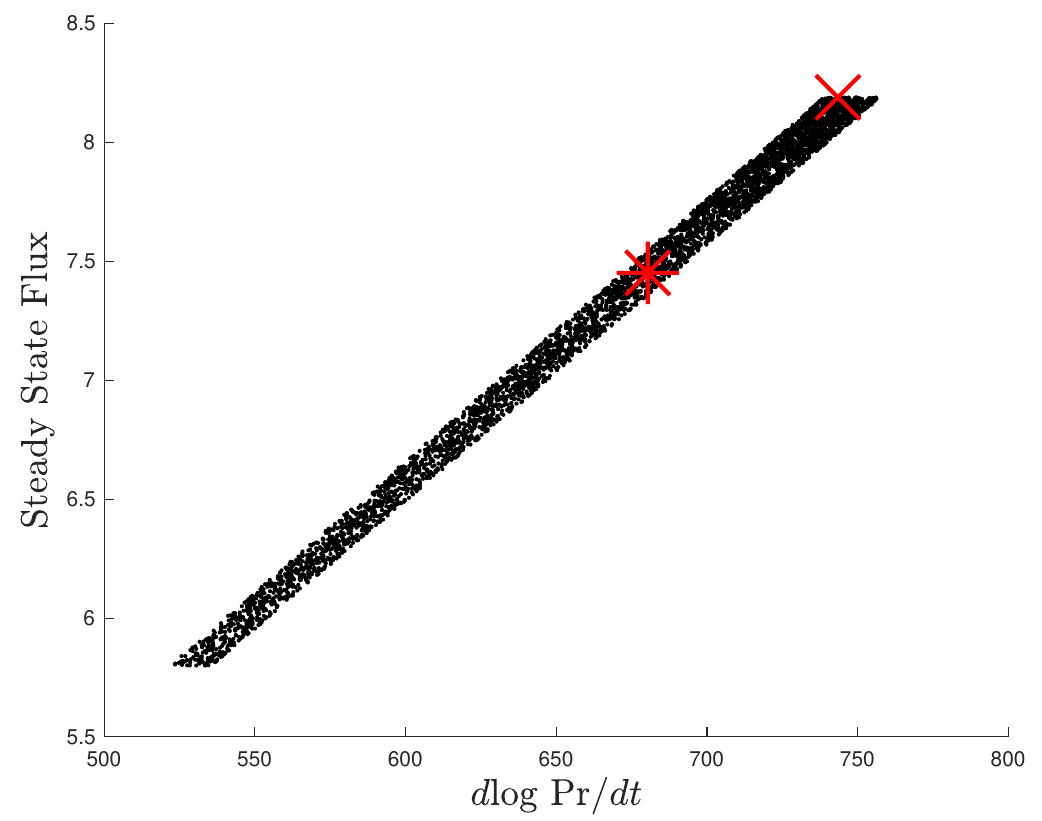}
\caption{(A) Free energy dissipation rate and (B) Entropy production rate are plotted against the steady state rate for each model in the ensemble generated by varying rate constants +/-2.0 orders of magnitude in a manner, described in the text, such that each model has a different steady state rate. The maximum entropy model, red $\ast$ at 680, 7.45) is compared to the random models (black dots). The highest entropy production rate (red X) occurs at the location 748, 8.19}
\label{fig:dissipation}
\end{figure}

A population of models can be generated that do not have the same steady state rates or free energy dissipation rates. This is done following the same process as before in which rate constants are determined by solving Eqn. \eqref{rate_constants1} using random steady state concentrations. However, in this case, the rate constant for the first reaction in the pathway, glucose kinase (HEX1), is set to the same value determined from the maximum entropy solution. Since the concentration of the product of that reaction, glucose 6-phosphate, is still randomly varied and the resulting steady state flux is solved using Eqn \ref{kinetic_law}, then if the latter concentration is less than its maximum  concentration, the reaction will be more favorable ($\Delta G_{\text{Hex1}}$ decreases), and the reverse reaction will occur less frequently. This process results in varying steady state and free energy dissipation rates, as shown in Figure \ref{fig:dissipation}A, and varying steady state and entropy production rates, as shown in Figure \ref{fig:dissipation}B. The entropy production rates (free energy dissipation) increase (decrease) nearly linearly with respect to increases in the steady state rate. The maximum entropy solution to the boundary value problem doesn't have the highest free energy dissipation or entropy production rate since it does not have the highest steady state rate, demonstrating Condition 2 in Section \ref{sec:most_prob_state}.

Since the maximum entropy model results in each reaction being equally far from equilibrium, there are fewer reactions close to equilibrium in the model. Consequently, that the maximum entropy models have generally high steady state rates is consistent with the concept that reactions near equilibrium in systems of coupled reactions will limit the rate of the overall pathway \cite{Noor2014}.

\subsection{\label{sec:level2}Probability of Models with High Dissipation Rates} 
\begin{figure}[hbt]
\centering
\includegraphics[width=1.0\linewidth]{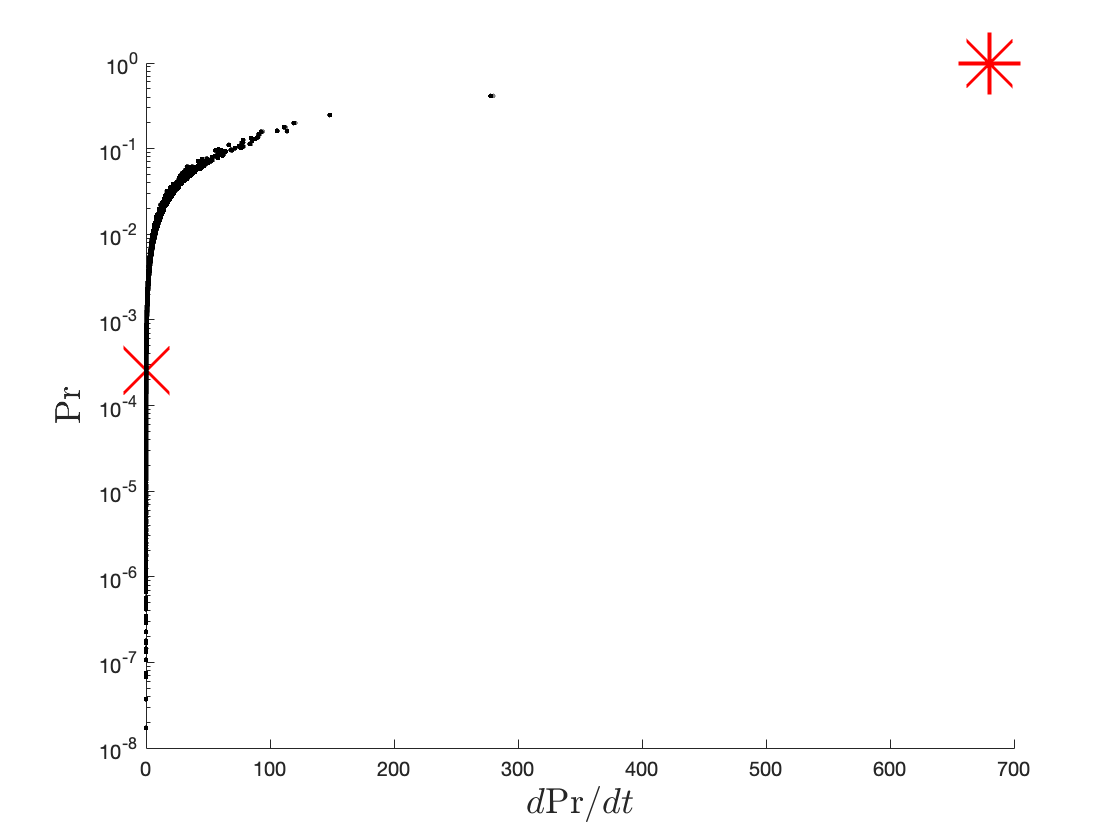}
\caption{ The relative probability Pr of each state is plotted against $d$Pr$/dt $ (Eqn. \ref{dPr_dt})
for the models shown in Figure \ref{fig:dissipation} (black dots). For the random population, the model with the highest entropy production rate ($d \text{log} Pr/dt = -dG/dt $) is shown by a red $X$. The maximum  entropy solution is the red $\ast$, which is also the state of maximum change in probability with time. (For a steady state, $d\text{Pr}/dt$ is a constant but is zero only for the equilibrium steady state.)}
\label{fig:dPrdt_vs_dPr}
\end{figure}

While it would seem from Figure \ref{fig:dissipation} that species with the highest steady state rate of metabolism might be the most optimal model due to correlation with the free energy dissipation rate, what is not taken into account in the analysis are the probabilities of each model relative to each other. In physical systems it is not the entropy production rate, $d \text{log} \text{Pr}/dt$, that is maximized but rather the change in probability of the system with time $d\text{Pr}/dt$.

To demonstrate, Figure \ref{fig:dPrdt_vs_dPr} plots the same models used in Figure \ref{fig:dissipation} (black dots) but this time comparing the probability of the steady state, $\text{Pr}$, with the change in probability with time, $d\text{Pr}/dt$. The probability of each of the steady states was determined using the Gaussian kernel method of Eqn \eqref{eq:empirical_rxn_pdf}, while $d\text{Pr}/dt$ was determined with 
Eqn. \eqref{dPr_dt}. \textcolor{black}{The probability densities for each of the 21 reactions are shown in Figure \ref{fig:rxn_density_distributions} in Appendix \ref{appdx:estimated_densities}.}

As can be seen in Figure \ref{fig:dPrdt_vs_dPr}, the maximum entropy model, indicated by the red $\ast$, has both the largest probability and the largest change in probability with time, $d\text{Pr}/dt$. That is, because of the principle of equiparition of energy, not only is the maximum entropy model expected to be the thermodynamically most likely model, it will also have the largest change in probability with time of any models. 
Unless one has specific information about a reaction not obeying the principle of equipartition of energy, perhaps because the catalyst can only lower its transition state a limited amount, the use of a maximum entropy model is the least biased model. A Michealis-Menton model, in contrast, is always biased by the assumption that there is infinite thermodynamic driving force to release the product of an enzymatic reaction, which is certainly not the case for coupled reactions in a cell. 

\subsection{\label{sec:level2}Prediction Variability due to Uncertainty in Standard Free Energies of Reaction.}\label{uncertainty1}
In order to determine the sensitivity of the maximum entropy predictions of fluxes and concentrations due to uncertainty in the estimates of $\Delta G^{\circ}$, we randomly sampled $\Delta G^{\circ}$ across its 95\% confidence interval and characterized the predictions of steady state fluxes and concentrations. In addition, each prediction was evaluated using multiple sets of random initial concentrations. The standard free energies come into play in the model predictions through the equilibrium constants, where the equilibrium constant for reaction $j$ is $K_j = e^{-\Delta G_j^{\circ}/RT}$. The collection of standard free energy changes, $\Delta G^{\circ}$ and standard deviations $\sigma (\Delta G^{\circ})$ for all reactions were calculated using eQuilibrator software \cite{Flamholz2012}.

\begin{figure}[hbt]
\centering
\includegraphics[width=1.0\linewidth]{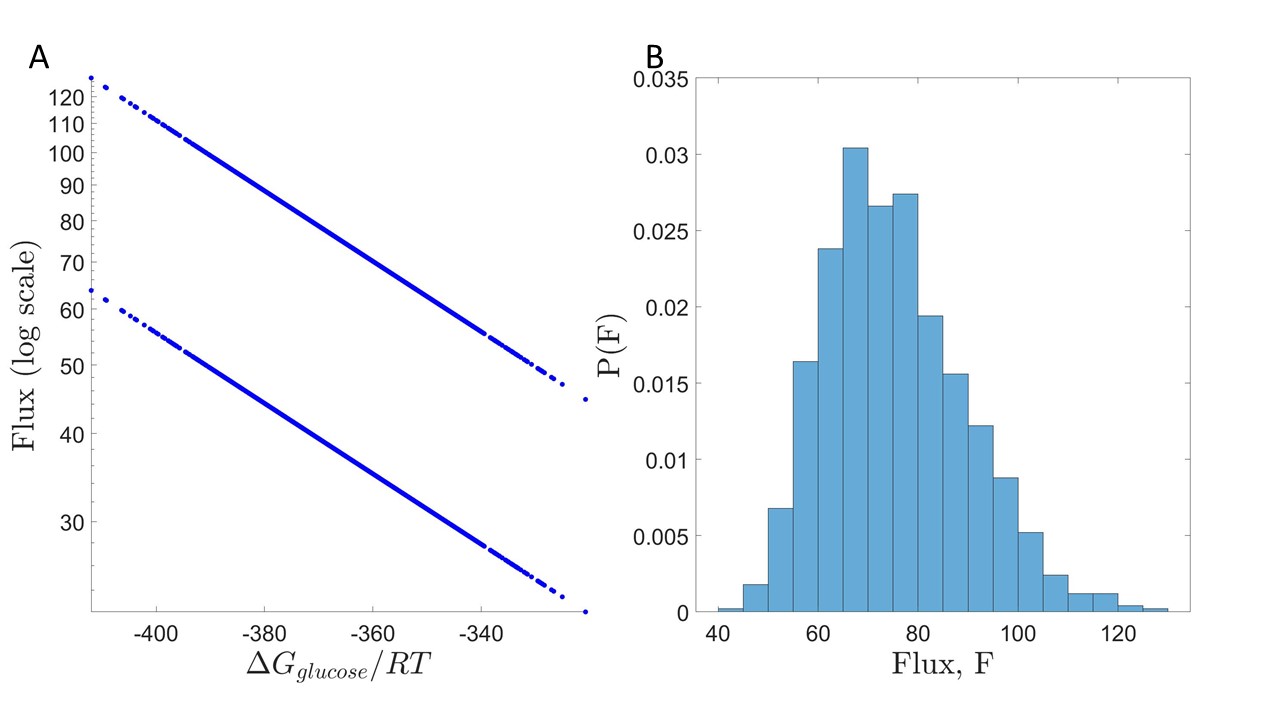}
\caption{(A) The relation between reaction affinity (Eqn. \eqref{delta_G_glucose2}) and steady state flux for the set of reactions from glucose through the TCA cycle and (B) corresponding probability distribution of the fluxes through upper glycolysis as a function of the uncertainty in the estimated standard free energies of reaction.}
\label{fig:keq_var_driving_fluxhist}
\end{figure}

 Specifically, 1000 values were sampled from disjoint intervals within a single standard deviation range, $ [\Delta G^{\circ} - \sigma (\Delta G^{\circ}), \Delta G^{\circ} + \sigma (\Delta G^{\circ})] $, using the Latin hypercube sampling (LHS) method \cite{Mckay1979,Marino2008}. The resulting equilibrium constants were calculated from the sampled free energy values. Ten initial starting concentrations were then chosen for each variable metabolite. Nonlinear least squares optimization using Eqns. \eqref{fluxes} and \eqref{optimization} was then used to acquire steady state fluxes through the reactions, metabolite concentrations and net driving forces on the reactions. The thermodynamic driving force on a reaction, known as the reaction affinity,  is defined as, 
\begin{equation}\label{reaction_affinity}
\begin{split}
    &{{{A}}_{j}}=-\frac{\partial G}{\partial {{\xi }_{j}}} \\ 
    &= \beta^{-1} \log {{K}_{j}}Q_{j}^{-1}.
\end{split}
\end{equation}
(If the system is at steady state and the concentrations are large enough that they can be assumed to be continuously distributed, then $-\Delta G_j \approx {A}_j$.) From the net driving force on each reaction, the total driving force on the pathway, $A$, was determined using Eqn. \eqref{reaction_affinity} and the relation $A=\sum_j A_j$. The variation of fluxes obtained as a function of the total driving force on the pathway, shown in Figure \ref{fig:keq_var_driving_fluxhist}A and Appendix \ref{app_uncertainty}, Table \ref{tab:variation_Flux_FreeEnergy}, results in a linear response. In the upper glycolysis pathway, the reactions produce exactly half the flux as downstream reactions due to the production of two glyceraldehyde-3-phophates from fructose-1,6-bisphophate in the fructose-bisphosphate aldolase catalyzed reaction. This results in a second line of higher flux data points for lower glycolysis (Figure \ref{fig:keq_var_driving_fluxhist}A). \textcolor{black}{The significance is that uncertainties in standard reaction free energies $\Delta G^\circ/RT$ can potentially have a significant effect on predicted reaction fluxes. This is the case whether one uses a maximum entropy model or not, since variation in the equilibrium constants $K_j$ results in variation in the rate parameters $k_j$ and $k_{-j}$, as well.}

The distribution of steady state fluxes from upper glycolysis is shown in the histogram in Figure \ref{fig:keq_var_driving_fluxhist}B. The mean flux was 75.5 and the standard deviation was 14.17. The maximum entropy model has a flux of 74.2 for upper glycolysis. 
Note that standard free energies are sampled uniformly across the 95\% confidence interval by virtue of the LHS method but the resulting variation in steady state flux has a bell-shaped distribution. \textcolor{black}{Consequently, even though uncertainties in $\Delta G^\circ$ can potentially result in a large range in steady state flux ([130,22], Figure \ref{fig:keq_var_driving_fluxhist}A), most of the variation in flux falls in a much smaller region of $75.5 \pm 14.2$.}

\begin{figure}[hbt]
\centering
\includegraphics[width=1.0\linewidth]{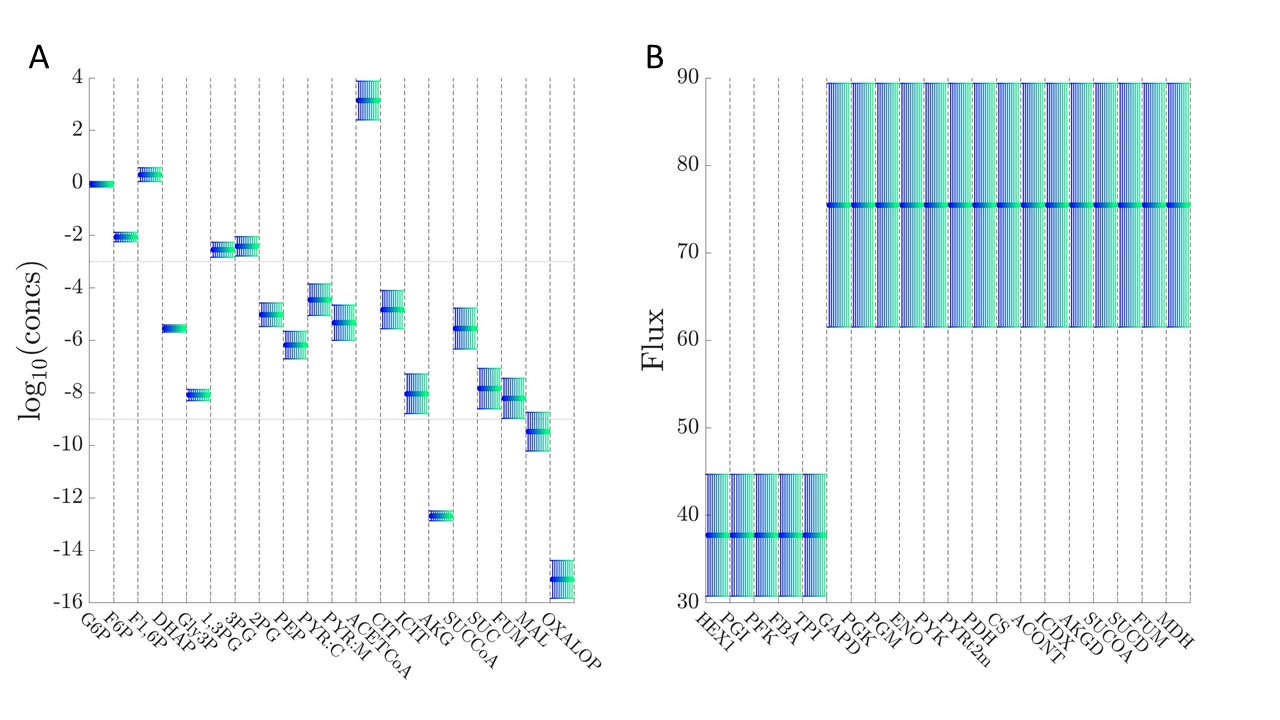}
\caption{Individual steady state metabolite levels (left) and fluxes (right) calculated via optimization from 10 unique starting concentrations are shown in blue-green in separate columns. Differences resulting from variation in the standard free energy changes of the reactions, $\Delta G^{\circ}$, are shown for each metabolite. Error bars represent a single standard deviation.}
\label{fig:keq_var_met_flux}
\end{figure}

Variation in steady state metabolite concentrations and individual reaction fluxes due to variability in standard reaction free energies are shown in Figure \ref{fig:keq_var_met_flux}. (The mean flux and standard deviations for individual reaction fluxes and free energies are again listed in Appendix \ref{app_uncertainty}, Table \ref{tab:variation_Flux_FreeEnergy}.) The variability in the predicted concentrations and flux is a result of variability in the equilibrium constants and not initial metabolite concentrations. To show this, we performed 10 distinct optimization routines with random initial starting concentrations for each metabolite. Each set of random starting metabolite concentrations was generated from a logarithmic scale using uniform distribution on the interval [-10, 0].  The mean and variation in concentrations due to uncertainty in $\Delta G^{\circ}$ are shown as error bars for each metabolite or reaction in Figure \ref{fig:keq_var_met_flux}A. The resulting steady state reaction flux for each of the 10 starting concentrations are represented by the range of colors from blue ($1^{st}$) to green ($10^{th}$). Identical means and standard deviations across each blue-green set of steady state values show that steady state metabolite and flux variation is not a result of starting concentrations or the optimization routine. Instead, steady state metabolite and flux variation is directly related to variations in the free energy. The large variability in the steady state flux in Figure \ref{fig:keq_var_met_flux}B is most likely due to large variability in standard reaction free energies of just a few reactions. In particular, Glyceraldehyde 3-phosphate dehydrogenase (GAPD) has an especially large coefficient of variation of $5.27\times 10^{4}$ (Appendix \ref{app_uncertainty}, Table \ref{tab:variation_Flux_FreeEnergy}), due to the small value of the equilibrium constant (mean $= 2.5\times 10^{-5}$, $\sigma = 1.34$). \textcolor{black}{The significance is that predicted concentrations do not vary over a large range due to uncertainties in the standard free energies of reaction $\Delta G_j^\circ$ used in the model.}

The steady state metabolite concentrations predicted from a maximum entropy approach or any other approach may be outside the range of the values typically observed experimentally \cite{Park2016, Bennett2009}, as can be seen for acetyl CoA and fructose-1,6-bisphosphate in Figure \ref{fig:keq_var_met_flux}. However, as we have shown elsewhere  \cite{Britton2020, King_frontiersSysBio_2023}, these metabolite concentrations are reduced to physiological levels when enzyme activities are introduced into the governing differential equations.



\subsection{\label{sec:level2}Prediction Uncertainty due to Variability in Ionic Strength}
The value of ionic strength, $I$, in the cell cytoplasm is often taken as $I=0.25$ M. In order fully test the model prediction sensitivity due to change in ionic strength, 1000 values were sampled uniformly within equally spaced disjoint intervals between [0.01, 0.5], which were chosen since they span the range of ionic strengths that are physically observed. Each value of ionic strength is then utilized to generate standard free energies for each reaction, $\Delta G_{j}^{\circ}(I)$, using eQuilibrator API \cite{Flamholz2012} which allows for the adjustment of free energies as a function of pH and ionic strength. For the overall chemical equation for the pathway, the standard free energy decreases as the ionic strength increases. As above, from standard free energies of reaction, equilibrium constants are calculated and subsequently used to obtain steady-state predictions for metabolites and fluxes.  

\begin{figure}[hbt]
\centering
\includegraphics[width=1.0\linewidth]{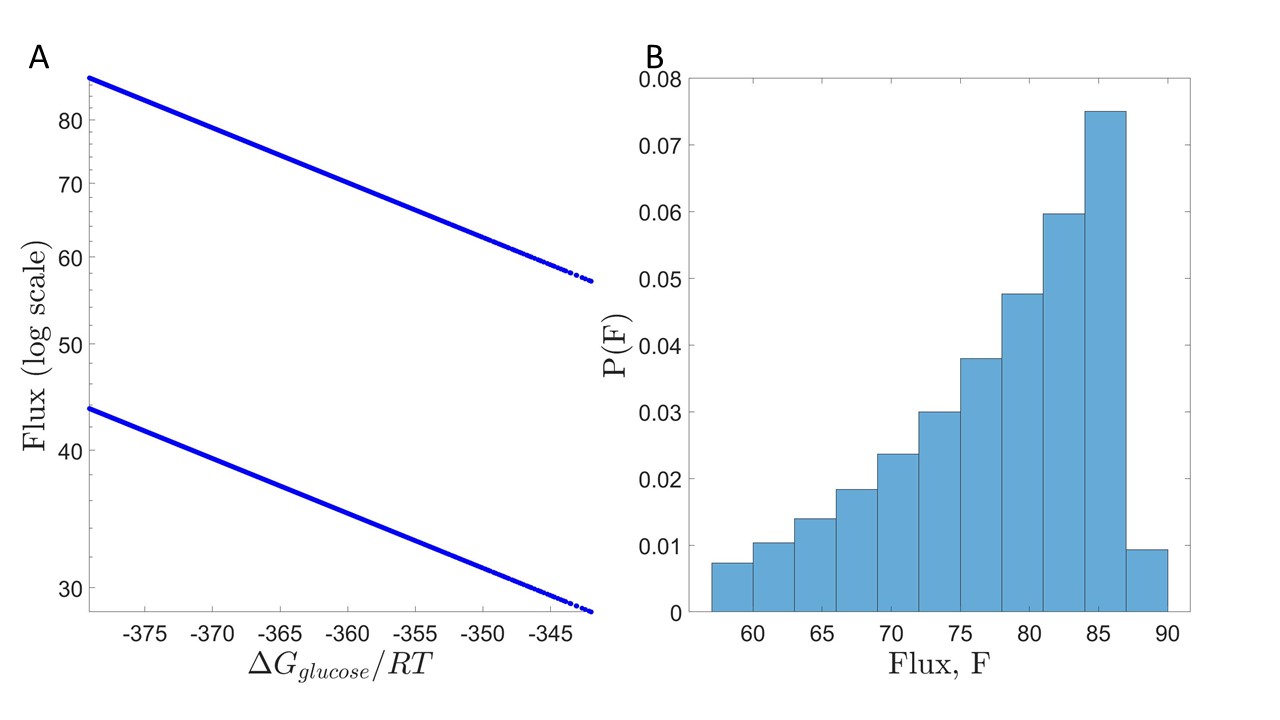}
\caption{The relation between reaction affinity and flux (left) is shown for the overall set of reactions from glucose through the TCA cycle as a function of the ionic strength used to determine the standard free energy changes of the reactions, $\Delta G^{\circ}$. The corresponding distribution of fluxes (right) are shown for the same ionic strength values. }
\label{fig:ionic_var_driving_fluxhist}
\end{figure}

\begin{figure}[hbt]
\centering
\includegraphics[width=1.0\linewidth]{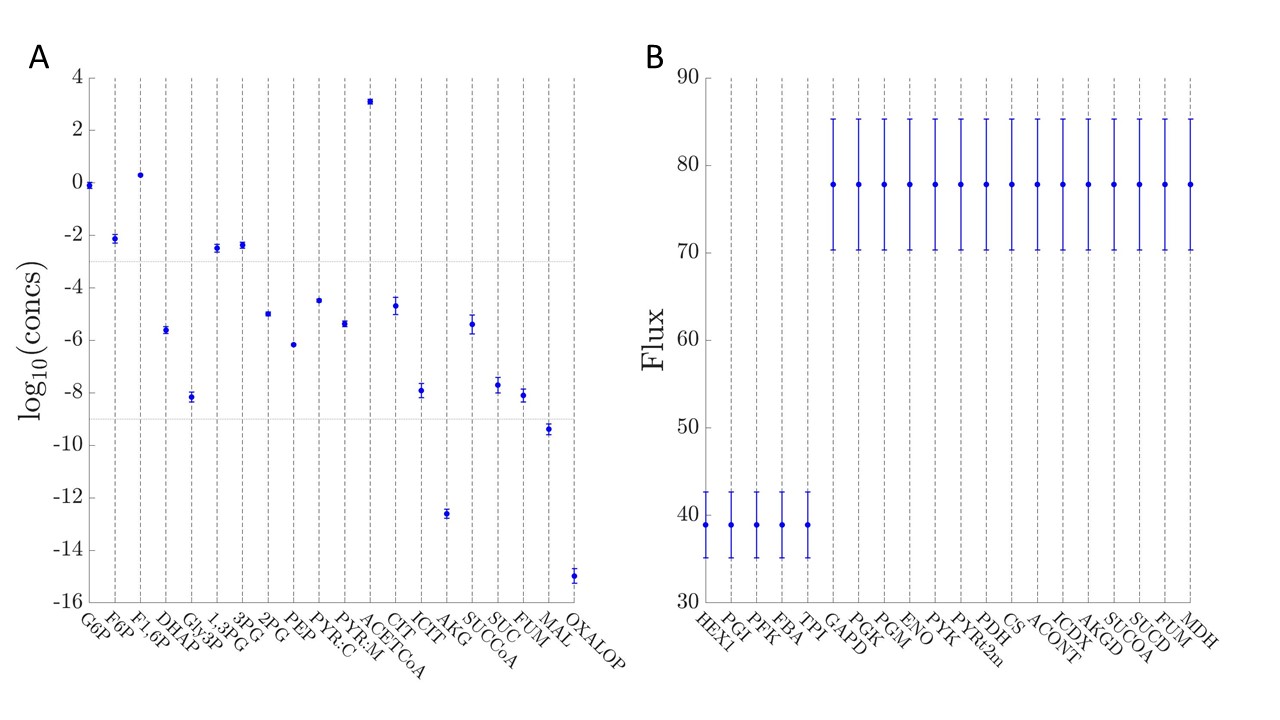}
\caption{Variability of the steady state metabolite concentrations (left) and fluxes (right) as a function of the ionic strength used to determine the standard free energy changes of the reactions, $\Delta G_{j}^{\circ}$.}
\label{fig:ionic_var_met_flux}
\end{figure}

The steady state fluxes due to variability in ionic strength in upper and lower glycolysis remain linear with respect to the reaction affinity, $A$ (Eqn. \eqref{reaction_affinity}), as shown in Figure \ref{fig:ionic_var_driving_fluxhist}A. This behavior matches what was previously observed when the free energy change for reactions was directly varied (Figure \ref{fig:keq_var_driving_fluxhist}A). However, the range of reaction affinities are reduced when ionic strength is varied, specifically the total affinity was within $[-379.225, -342.274]$ \textcolor{black}{(Figure \ref{fig:ionic_var_driving_fluxhist}A)}, but ranged between $[-413.483 , -316,46]$ due to uncertainty in equilibrium constants \textcolor{black}{(Figure \ref{fig:keq_var_driving_fluxhist}A). Consequently, obtaining accurate standard free energies of reaction is more important than getting accurate ionic strength values for inside the cell.} Moreover, the distribution of fluxes through upper glycolysis (Figure \ref{fig:ionic_var_driving_fluxhist}B) is no longer bell-shaped as was observed in Figure \ref{fig:keq_var_driving_fluxhist}B, due to the distribution of standard free energies derived from the sampled ionic strength values. 

Variation in individual steady state metabolite concentrations due to alterations in ionic strength (Figure \ref{fig:ionic_var_met_flux}A) are represented by blue error-bars showing a single standard deviation. Similarly, the steady state flux through each reaction due to alterations in ionic strength are shown in Figure \ref{fig:ionic_var_met_flux}B. In both the steady state metabolite concentrations and reaction flux, the variations due to ionic strength lie within the variations based on uncertainty (95\% confidence interval) in the standard free energy change (compare Figure \ref{fig:keq_var_met_flux}B and Figure \ref{fig:ionic_var_met_flux}B). The mean flux and standard deviations for individual reaction fluxes and free energies are listed in Appendix \ref{app_uncertainty}, Table \ref{tab:variation_Flux_FreeEnergy}. \textcolor{black}{Like variations in flux, the variation in concentrations due to uncertainties in ionic strength are much less that the variations seen due to uncertainty in standard reaction free energies, $\Delta G_j^\circ$.}

\section{\label{sec:level1}Discussion and Conclusion}

In this work, the relationship between the probability density of a system of coupled chemical reactions, with each chemical species represented by it’s Boltzmann probability, and the rate of free energy dissipation, entropy production, and change in probability with time, $d\text{Pr}/dt$, due to chemical reactions was demonstrated. 
When the total number of particles does not change in any reaction, the free energy dissipation rate is the same as the entropy production rate, but opposite in sign. The entropy production rate for systems in which the total number of particles can vary will additionally depend on changes in the probability density due to changes in the number of particles. We showed that, while the maximum entropy state is not necessarily the state with the highest entropy production rate ($d\log\text{Pr}/dt$), it is the state that maximizes the change in probability with time, $d\text{Pr}/dt$. 

The entropy production rate relates the chemical master equation to chemical kinetics, as described by the law of mass action.  Moreover, the time-dependent probability density for a system of coupled chemical reactions can be separated into time-dependent and time-independent equations (Eqn \eqref{kinetic_law_expanded}), which can be solved separately by first solving for the maximum entropy solution and then adding specific time dependence back into the rate equations using either known rate constants or measured steady state concentrations.

Predictions between maximum entropy models and kinetic models which do not maximize the entropy production were compared. Systems that have the same steady state rate but only differ in free energies at each reaction have the same free energy dissipation rate, regardless. Only when the steady state rates vary, do the free energy dissipation rates also vary. 

In a maximum entropy model, of course, free energies are equally partitioned across reactions as much as possible. The equipartition of reaction energy acts to generally, but not always, increase the steady state rate relative to other models (i.e., models that are not at maximum entropy) because fewer reactions in the maximum entropy model are near equilibrium. Reactions near equilibrium act to constrain the steady state flux \cite{Noor2014}. The closer to  equilibrium, the relatively more flux occurs in the reverse direction. 
Despite this, a maximum entropy solution is not guaranteed to have the highest nominal rate of entropy production or free energy dissipation.  

However, a maximum entropy model is also maximal in probability space, and consequently when the probabilities of each model are considered, the maximum entropy model strikes a balance between a high free energy dissipation and entropy production rates and the thermodynamic work required to maintain the steady state. Because of the link between free energy and probability density (Eqn. \eqref{probability_general}), a low probability density means that a significant amount of energy is required to obtain that steady state.  

In contrast, maximum entropy models are highly probable because they represent the thermodynamic average of a population. If a population average model can’t provide a description of a biological process of interest, it is likely that the model is incomplete and missing key functionality. Consequently, if a maximum entropy model does not adequately capture the experimentally observed behavior, the model is likely missing important mechanisms or functionality. In this regard, it is essential to consider physical, chemical and biological constraints, as well, for these may represent the missing, and possibly emergent, functionality. 
For instance, since the concentration of chemical species in maximum entropy models are proportional to their Boltzmann probabilities, adjusted for non-equilibrium boundary conditions, the concentrations for species such as acetyl CoA and fructose-1,6-bisphosphate may be unreasonably high for a biological system. However, application of constraints in the form of activity coefficients for the system results in concentrations that are consistent with experimental observation, and hence represents an approach to learn regulation \cite{Britton2020,King_frontiersSysBio_2023}.

The metabolic models shown in this work are reduced models that are appropriate when the details of the dynamics of the catalytic process are not needed. That is, the models represent dynamics that are coarse grained over the enzyme binding and release of substrates and products, and consequently can be solved quickly using non-linear least squares optimization of the metabolite concentrations to steady state values. In principle, maximum entropy simulation models can be developed analogously at the enzyme level by taking into account the thermodynamic work required to synthesize the enzymes and the benefit provided to the cell \cite{Sivak2014}. Maximum entropy methods have been developed to replicate experimental observations based on data \cite{Firman2017}.

Relative errors in the free energies of reaction do not have a major impact on the maximum entropy solution. The predicted concentrations may vary within $\pm 1-3$ orders of magnitude. For comparison, concentrations in an \textit{E. coli}  cell can vary by more than 7 orders of magnitude, from tens of millimolar to approximately nanomolar which corresponds to 1 molecule per cell. 

However, the relative uncertainty in standard free energies can have an impact on steady state flux by up to 70\%.  Still, as models become more precise and accurate it will be important to reduce the uncertainties in standard free energies of reaction as much as possible. Sophisticated electronic structure calculations are perhaps the most tractable way to develop more accurate models \cite{Bylaska2006, Jinich2014}.

Likewise, variation in ionic strength not have a strong effect on concentrations, but may impact steady state fluxes more significantly since the fluxes are a function of the overall driving force on the pathway, and the overall driving force increases with an increases in the ionic strength. Impacts on steady state flux up observed to be as high as 27\%.

\begin{acknowledgments}
W. R. Cannon was supported by the DOE Office of Biological and Environmental Research, through projects 74860. S. Britton was supported by a U. S. Department of Energy (DOE), Office of Science Graduate Student Research award and by funding from the DOE Office of Biological and Environmental Research, through project 74860. M. Alber was supported by funding from the National Science Foundation, Grant DMS-1762063, through the joint NSF DMS/NIH NIGMS Initiative to Support Research at the Interface of the Biological and Mathematical Sciences. PNNL is operated by Battelle for the US Department of Energy under Contract DE-AC06-76RLO. 
\end{acknowledgments}

\section*{AUTHOR DECLARATIONS}
\subsection*{Author Contributions}
W.R. Cannon developed the probability and entropy theory and application to mass action models. S. Britton and M. Banwarth-Kuhn designed the uncertainty quantification analysis due to variation in free energies. M. Alber advised in the design of the uncertainty quantification methods. All authors contributed to writing the article.

\appendix

\section{\label{app_oddsRatio}Odds Ratio of States}
Consider states $J$ and $J+ \delta \xi_j$ in which
$\delta \xi_j$ is the extent of reaction $j$ such that $\delta \xi_j = 0, 1, 2,..$ correpsonds to zero, one, two or more firings of reaction $j$ transforming the reactants in state $J$ with concentrations $n_i(J)$ into products in state $J+\delta \xi_j$ with concentrations $n_i(J+ \delta \xi_j)$. From Eqn \ref{probability_general} in the main text, the ratio of the probability density functions is,

\begin{eqnarray}
\frac{Pr(J+\delta\xi_j)}{Pr(J)} & = & \frac{N(J+\delta \xi_j)!}{N(J)!} 
    \cdot \\ 
    && \prod_i^{M} \frac{n_i(J)!}{n_i(J+ \delta \xi_j) !}
    \cdot 
    \theta_i^{n_i(J + \delta \xi_j)-n_i(J)} \nonumber 
\end{eqnarray}
We will assume that $N(J) = N(J + \delta \xi_j)$ such that the factorials over total counts $N$ cancels out. The concentrations $n(J + \delta \xi_j) = n(J) + \delta \xi_j \gamma_{i,j}$, where $\gamma_{i,j}$ is the signed stoichiometric coefficient for metabolite $i$ in reaction $j$. Assuming that $\delta \xi_j = 1$ and substituting,
\begin{eqnarray}
\frac{Pr(J+1)}{Pr(J)} & = & \prod_i^{M} \frac{n_i(J) !}{(n_i(J) +  \gamma_{i,j}) !}
    \cdot
    \theta_i^{\gamma_{i,j}}
    \label{sup:odds_ratio}
\end{eqnarray}
Using the unsigned stoichiometric coefficients $\nu_{i,j} = |\gamma_{i,j}|$, then depending on whether $\gamma_{i,j}$ is positive (products) or negative (reactants), the factorials for each metabolite $i$ is either a rising factorial or a falling factorial such that for products ($\gamma_{i,j} > 0$),
\begin{equation}
    \frac{n_i(J) !}{(n_i(J) + \gamma_{i,j}) !} = \frac{1}{(n_i + \nu_{i,j})\cdots(n_i + 1)} \approx n_i^{-\gamma_{i,j}},
\end{equation}
and for reactants ($\gamma_{i,j}  < 0$),
\begin{equation}
    \frac{n_i(J) !}{(n_i(J) +  \gamma_{i,j}) !} = {(n_i)(n_i - 1)\cdots(n_i - \nu_{i,j}+1)} \approx n_i^{-\gamma_{i,j}},
\end{equation}
where the approximations are valid when $n_i \gg 0$.
Eqn \ref{sup:odds_ratio} can then be written as,
\begin{eqnarray}
\frac{Pr(J+\delta\xi_j)}{Pr(J)} & 
    \approx & \prod_{i} n_i^{-\gamma_{i,j}}
    \cdot
    \theta_i^{\gamma_{i,j}} \\
    & = & K_j Q^{-1}_j
\end{eqnarray}
in which $K_j = \prod_i \theta_i^{\gamma_{i,j}}$ and $Q_j = \prod_i n_i^{\gamma_{i,j}}$.
\\
\section{\label{app_uncertainty}Uncertainty and Variability}
\subsection{\label{app_concs}Concentrations in Figure 2D}
The metabolite concentrations for feasible reactions in Figure 2D vary $\pm$ 1.0 order of magnitude from the maximum entropy solution, as expected since the range of sampled metabolites was 2.0 orders of magnitude centered on the maximum entropy solution. The exception is $\alpha$-ketoglutarate (AKG), for which only values up to +1.0 order of magnitude greater than the maximum entropy solution were found. This is likely due to $\alpha$-ketoglutarate also being the product of the GOGAT reaction (not shown), which is an input boundary reaction that operates at equilibrium in the maximum entropy model. While the models operate fine without the GOGAT reaction and the observed free energies fall into the same distributions, the metabolite concentrations of the TCA cycle are under-determined without the GOGAT reaction. This is because the TCA cycle consists of eight reactions but nine variable metabolites because the entry point into the cycle is the citrate synthase reaction, which has both Acetyl co-enzyme A and oxaloacetate as free variables. For comparison, in each box in Figure \ref{fig:variation_rates}D a black $\ast$ denotes values for the maximum entropy solution while a red $\ast$ denotes values of the distribution that is furthest away from the maximum entropy distribution regarding reaction free energies, and a blue $\ast$ denotes values of the distribution that is furthest away from the maximum entropy distribution regarding reaction rates.

\subsection{Variability in Flux and Reaction Free Energies}
Table \ref{tab:variation_Flux_FreeEnergy} shows the sampling statistics due to the Latin hypercube sampling associated with determining the variability of reaction fluxes and free energies (1) associated with  the uncertainty in the standard free energies of reaction (at an ionic strength $I= 0.25)$ and (2) associated with variability in the standard free energy of reaction across a range of ionic strengths, $I \in [0.01, 0.50]$. 

\begin{widetext}
\begin{table*}
\caption{Statistics characterizing the variability in reaction fluxes and standard reaction free energies due uncertainty in standard reaction free energies $\Delta G^{\circ}(I)$ at $I=0.25$, and variability of mean standard reaction free energies $\Delta G^{\circ}(I)$ due to differing ionic strengths $I$. For sampling standard reaction free energies at $I=0.25$, 1000 values were sampled from disjoint intervals within a single standard deviation range, $ [\Delta G^{\circ} - \sigma (\Delta G^{\circ}), \Delta G^{\circ} + \sigma (\Delta G^{\circ})]$ (95\% Confidence Interval, CI), using the Latin hypercube sampling (LHS) method \cite{Mckay1979,Marino2008}. For sensitivity of mean standard reaction free energies $\Delta G^\circ(I)$ due to different ionic strengths $I$, 1000 values were sampled uniformly within equally spaced disjoint intervals  [0.01, 0.5] of ionic strengths.}
\begin{ruledtabular}
\small
\begin{tabular}{lrrrrrrr|rrrrrrr}
\multicolumn{2}{c}{} &\multicolumn{6}{c|}{Variation in Flux, $\dot{\xi_j}$} &\multicolumn{6}{c}{Variation in Reaction $\Delta G^\circ$} \\
\multicolumn{2}{c}{} &\multicolumn{3}{c}{ \underline{95\%CI $\Delta G^\circ(I=0.25)$}} &\multicolumn{3}{c|}{\underline{$\Delta G^\circ(I), I \in[0.01,0.5]$}} &\multicolumn{3}{c}{\underline{95\%CI $\Delta G^\circ(I=0.25)$}} &\multicolumn{3}{c}{\underline{$\Delta G^\circ(I), I \in[0.01,0.5]$}} \\
\multicolumn{2}{c}{Sys./Reaction} &Mean &STD &CV &Mean &STD &CV &Mean\footnote{These are also the $\Delta G^\circ$ values used in the metabolic model for Figures \ref{fig:histogram}-\ref{fig:dPrdt_vs_dPr}.} &STD &CV &Mean &STD &CV \\
\hline
\multirow{12}{*}{Glycolysis} 
&HEX1 &37.76 &7.09 &0.19 &38.91 &3.75 &0.10 &-17.06 &0.41 &0.024 &-16.87 &0.40 &0.02 \\
&PGI &37.76 &7.09 &0.19 &38.91 &3.75 &0.10 &2.53 &0.34 &0.14 &2.52 &0.00 &0.00 \\
&PFK &37.76 &7.09 &0.19 &38.91 &3.75 &0.10 &-15.45 &0.51 &0.03 &-15.86 &0.98 &0.06 \\
&FBA &37.76 &7.09 &0.19 &38.91 &3.75 &0.10 &20.51 &0.50 &0.025 &21.10 &1.40 &0.07 \\
&TPI &37.76 &7.09 &0.19 &38.91 &3.75 &0.10 &5.49 &0.44 &0.08 &5.49 &0.02 &0.00 \\
&GAPD &75.52 &14.17 &0.19 &77.82 &7.50 &0.10 &6.69 &0.52 &0.08 &5.79 &2.16 &0.37 \\
&PGK &75.52 &14.17 &0.19 &77.82 &7.50 &0.10 &-18.47 &0.51 &0.028 &-18.50 &0.04 &0.00 \\
&PGM &75.52 &14.17 &0.19 &77.82 &7.50 &0.10 &4.20 &0.38 &0.09 &4.19 &0.04 &0.01 \\
&ENO &75.52 &14.17 &0.19 &77.82 &7.50 &0.10 &-4.08 &0.42 &0.10 &-4.08 &0.00 &0.00 \\
&PYK &75.52 &14.17 &0.19 &77.82 &7.50 &0.10 &-27.54 &0.54 &0.02 &-27.43 &0.29 &0.01 \\
&PYRt2m &75.52 &14.17 &0.19 &77.82 &7.50 &0.10 &-5.71 &0.00 &0.00 &-345.51 &0.33 &0.00 \\
&PDH &75.52 &14.17 &0.19 &77.82 &7.50 &0.10 &-43.92 &4.43 &0.10 &-44.04 &0.32 &0.01 \\
\hline
\multirow{9}{*}{TCA} 
&CS &75.52 &14.17 &0.19 &77.82 &7.50 &0.10 &-35.12 &0.54 &0.015 &-35.56 &1.02 &0.029 \\
&ACONT &75.52 &14.17 &0.19 &77.82 &7.50 &0.10 &7.63 &0.42 &0.06 &7.63 &0.00 &0.00 \\
&ICDH &75.52 &14.17 &0.19 &77.82 &7.50 &0.10 &-2.87 &4.40 &1.53 &-2.73 &0.33 &0.12 \\
&AKDG &75.52 &14.17 &0.19 &77.82 &7.50 &0.10 &-36.35 &4.60 &0.13 &-36.90 &1.33 &0.04 \\
&SUCCoA &75.52 &14.17 &0.19 &77.82 &7.50 &0.10 &1.92 &0.86 &0.44 &1.97 &0.16 &0.08 \\
&SUCD &75.52 &14.17 &0.19 &77.82 &7.50 &0.10 &0.00 &1.34 &52748.35 &62.54 &0.68 &0.01 \\
&FUM &75.52 &14.17 &0.19 &77.82 &7.50 &0.10 &-3.45 &0.35 &0.10 &-3.45 &0.00 &0.00 \\
&MDH &75.52 &14.17 &0.19 &77.82 &7.50 &0.10 &29.99 &0.24 &0.01 &29.71 &0.68 &0.02 \\
&GOGAT & 0.00 &0.00 &-6.14 &0.00 &0.00 &-2.96 &48.90 &1.04 &0.021 &48.44 &1.00 &0.02 \\
\bottomrule
\end{tabular}
\end{ruledtabular}
\label{tab:variation_Flux_FreeEnergy}
\end{table*}
\end{widetext}

\begin{widetext}
\subsection{\label{appdx:estimated_densities}Estimated Probability Densities for Reactions}

\begin{figure}
    \centering
    \includegraphics[width=1.0\linewidth]{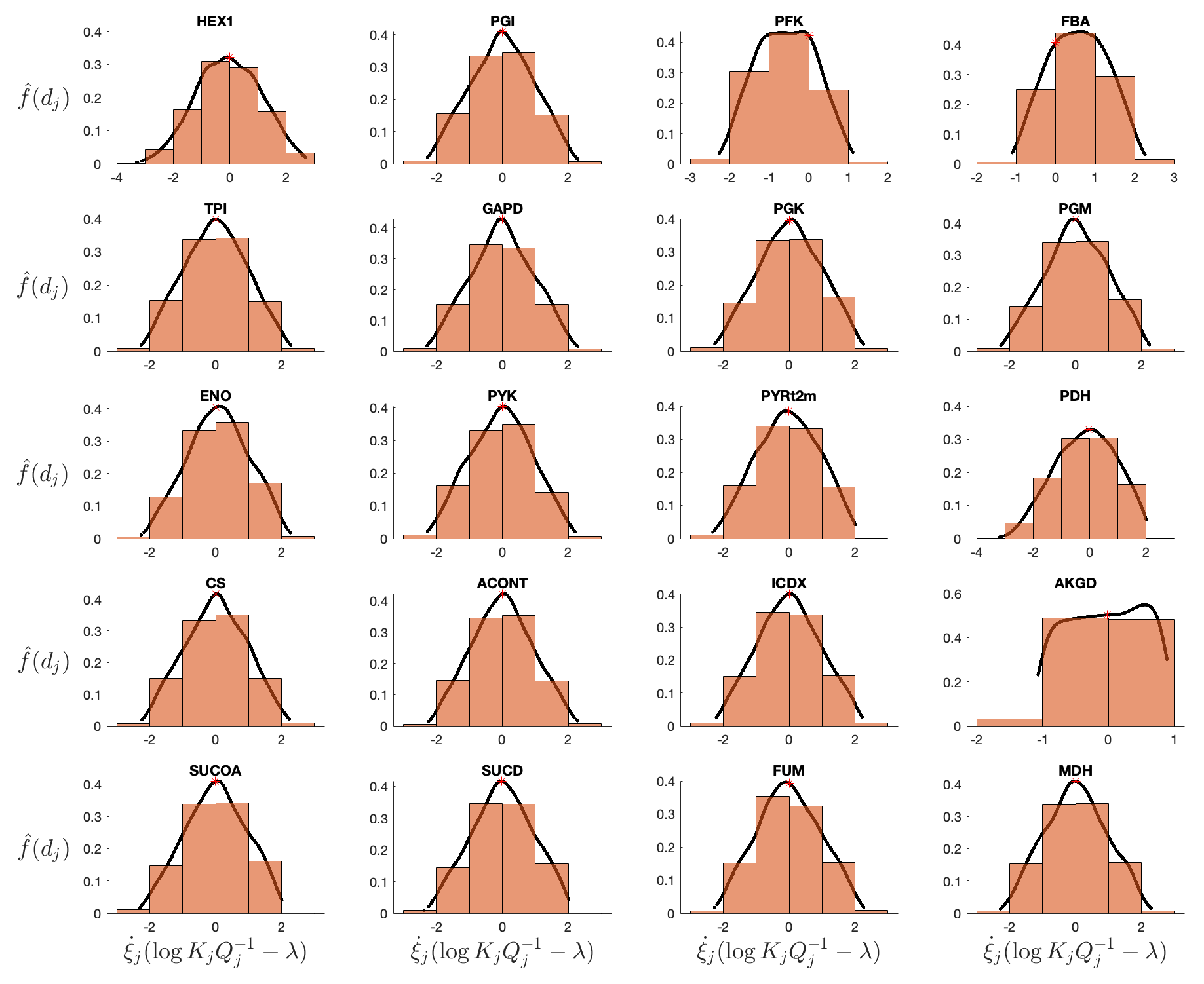}
    \caption{Estimated probability densities $\hat{f}(d_{j})$ and histograms for the distance $d_{j} = \log K_jQ_j^{-1} - \lambda$ from the most probable density for each of the 20 reactions used in the metabolic model that have reaction flux $\dot{\xi}_j \neq 0$. Each density is calculated from Eqn \ref{eq:empirical_rxn_pdf_partial}. The true maximum entropy solution occurs at $d_j = 0$, the Marcelin approximation to the maximum entropy solution is indicated by a red $\ast$ and the location of the peak density in each distribution occurs at the average value of $d_j$ from the distribution of Eqn \eqref{eq:empirical_rxn_pdf_partial}.  See section \ref{sec:empirical_prob} for details.}
    \label{fig:rxn_density_distributions}
\end{figure}
\end{widetext}
\clearpage

\bibliography{library_regulation_paper}

\end{document}